\documentclass[lettersize,journal]{IEEEtran}
\ifCLASSOPTIONcompsoc
  \usepackage[nocompress]{cite}
\else
  \usepackage{cite}
\fi

\ifCLASSINFOpdf
\else
\fi
\newcommand{\etal}{\textit{et al. }}
\usepackage{balance}
\usepackage{breqn}
\usepackage{array}
\usepackage{colortbl,booktabs}
\usepackage{subcaption}
\usepackage{titlesec}
\usepackage{algorithmic}
\usepackage{algorithm}
\usepackage{dblfloatfix}
\usepackage{mathtools,amssymb}
\usepackage{tabularx}
\usepackage{pgfplots} 
\usetikzlibrary{patterns}
\usepackage{graphicx}
\begin{document}
\title{\textit{Trust--SIoT}: Towards Trustworthy Object Classification in the Social Internet of Things}

%

% \author{Shell,~\IEEEmembership{Member,~IEEE,}
%         John~Doe,~\IEEEmembership{Fellow,~OSA,}
%          and~Jane~Doe,~\IEEEmembership{Life~Fellow,~IEEE}% <-this % stops a space
% \IEEEcompsocitemizethanks{\IEEEcompsocthanksitem M. Shell was with the Department
% of Electrical and Computer Engineering, Georgia Institute of Technology, Atlanta,
% GA, 30332.\protect\\
% % note need leading \protect in front of \\ to get a newline within \thanks as
% % \\ is fragile and will error, could use \hfil\break instead.
% E-mail: see http://www.michaelshell.org/contact.html
% \IEEEcompsocthanksitem J. Doe and J. Doe are with Anonymous University.}}% <-this % stops a space
% % \thanks{Manuscript received April 19, 2005; revised August 26, 2015.}}

\author{Subhash~Sagar,
        Adnan~Mahmood,
        Kai Wang,
        Quan~Z.~Sheng,
        Jitander~Kumar~Pabani,
        and~Wei~Emma~Zhang% <-this % stops a space
\thanks{Subhash Sagar, Adnan Mahmood, Kai Wang, and Quan Z. Sheng are with the School of Computing, Macquarie University, Sydney,
NSW 2109, Australia, e-mail: \{subhash.sagar, adnan.mahmood, kai.wang, michael.sheng\}@mq.edu.au.}
\thanks{Jitander Kumar is with the School of Telecommunication Engineering, University of Malaga, Malaga, 29016, Spain, e-mail: jitander.pabani@uma.es.}
\thanks{Wei Emma Zhang is with the School of Computer Science, 
%Faculty of Engineering, Computer, and Mathematical Sciences, 
The University of Adelaide, Adelaide, SA 5005, Australia, e-mail: wei.e.zhang@adelaide.edu.au.}
}

\markboth{IEEE Transactions on Network and Service Management,~Vol.~, No.~, April~2022}%
{Shell \MakeLowercase{\textit{et al.}}: Bare Advanced Demo of IEEEtran.cls for IEEE Computer Society Journals}

\IEEEtitleabstractindextext{%
\begin{abstract}

%Adnan: @Subhash - I have now Edited the Abstract. Please check.

The recent emergence of the promising paradigm of the Social Internet of Things (SIoT) is a result of an intelligent amalgamation of the social networking concepts with the Internet of Things (IoT) objects (also referred to as ``things'') in an attempt to unravel the challenges of network discovery, navigability, and service composition. This is realized by facilitating the IoT objects to socialize with one another, i.e., similar to the social interactions amongst the human beings. A fundamental issue that mandates careful attention is to thus establish, and over time, maintain trustworthy relationships amongst these IoT objects. Therefore, a trust framework for SIoT must include object-object interactions, the aspects of social relationships, credible recommendations, etc., however, the existing literature has only focused on some aspects of trust by primarily relying on the conventional approaches that govern linear relationships between input and output. 
%Therefore, 
In this paper, an artificial neural network-based trust framework, \emph{Trust--SIoT}, has been envisaged for identifying the complex non-linear relationships between input and output in a bid to classify the trustworthy objects. Moreover, \emph{Trust--SIoT} has been designed for capturing a number of key trust metrics as input, i.e., direct trust by integrating both current and past interactions, reliability and benevolence of an object, credible recommendations, and the degree of relationship by employing knowledge graph embedding. Finally, we have performed extensive experiments to evaluate the performance of \emph{Trust--SIoT} vis-\'a-vis state-of-the-art heuristics on two real-world datasets. The results demonstrate that 
%the 
\emph{Trust--SIoT} achieves a higher F1 
%score with 
and 
lower MAE and MSE scores.

\end{abstract}

}

\maketitle

% conference papers position the abstract like regular (non-compsoc)
\IEEEdisplaynontitleabstractindextext
% \IEEEdisplaynontitleabstractindextext has no effect 
% % creates the second title. It will be ignored for other modes.
\IEEEpeerreviewmaketitle

\section{Introduction}
\label{sec:introduction}

\IEEEPARstart{R}{ecent} advancements in computing technologies have witnessed a massive number of smart objects (e.g., smart meters, smart watches, and smart refrigerators) connected to the Internet to form the Internet of Things (IoT) \cite{8334540,CACM2021,ZhangSMTZHAASM20}. Furthermore, these smart objects are equipped with sensing, processing, and communication capabilities, allowing them to provide a variety of applications and services, which are expanding in a variety of areas, including personal, industrial, and commercial domains \cite{9427249} \cite{sagar_survey}.  
With the expansion of IoT applications ranging from smart homes,
%to 
smart factories. 
%to 
smart cities. 
%and 
to e-Health, the number of IoT objects (i.e., devices) are rapidly increasing, limiting network discovery and navigability when IoT devices consume and use services from one another. Over the last decade, several research have have investigated the concept of incorporating social interactions components with IoT devices to enhance information discovery in the same way that humans do using the principle of small-world phenomenon, which refers to the short chain of links (i.e., relationships) among individuals in societies \cite{nitti2014_n}\cite{10.5555/2980539.2980596}.

\begin{figure}[t]
    \centering
    \includegraphics[scale=0.5]{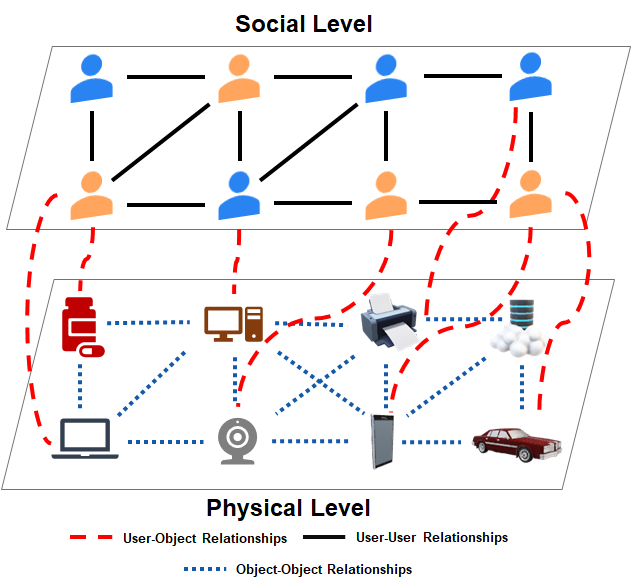}
    \caption{Depicting the SIoT relationships}
    \label{fig:siot_rel}
    \vspace{-1em}
\end{figure}

The mapping of the social structure of human and physical (i.e., IoT) devices (see Figure \ref{fig:siot_rel}) has led to an emerging paradigm of \emph{Social Internet of Things (SIoT)} and has opened the ways for next generation of IoT \cite{Atzori2011SIoTGA}. Furthermore, the mapping encompasses three distinct relationships (i.e., user-object relationships, object-object relationships, and user-user relationships). In SIoT, objects have the potential to socialize by establishing social relationships with one another autonomously based on the rules defined by their individual owners \cite{ATZORI20123594}. The evolution of SIoT can be foreseen as trillions of objects acting as autonomous agents (i.e., requesting and providing services) with a number of benefits, including but not limited to the assurance of effective object and information discovery in a trust-oriented manner, network scalability similar to human beings, building trustworthiness by incorporating the interaction behaviour among friends (or objects), and to utilize and extends existing social network models for SIoT networks.

Through the use of social relationships with IoT objects, SIoT has opened the way for the next generation of IoT. Maintaining trustworthy relationships, as well as security, privacy, and trust-related concerns, may however, limit the significance of SIoT paradigm. In the SIoT scenario, for example, a service requester (or \emph{trustor}) has the responsibility of observing the trustworthiness of the service providers (or \emph{trustee}) who provide the requested service; however, the possibility of a misbehaving object providing false services or false recommendations about a trustee to gain an advantage for a set of services can disrupt the availability and integrity of the SIoT services. Although some researches have presented cryptographic and non-cryptographic solutions to solve such concerns \cite{9011598} \cite{9382796}, security issues like 
%as 
trust as well as reputation are difficult to handle with such solutions. As a result, an effective SIoT trustworthiness management system is required to cope with misbehaving SIoT objects by limiting their services and by selecting only the credible and trustworthy objects before relying on the services provided by them.

The motive for establishing trustworthiness management for SIoT is evident based on the observations made. Several studies have been proposed to address the challenges of trustworthiness management. However, most of these studies do not consider a thorough study of SIoT fundamentals, such as 1) the selection of SIoT-based trust metrics, 2) only considering the static characteristics of SIoT relationships (e.g., ownership, location), and 3) using the traditional approach (i.e., linear relationship) to map the input and output for trustworthiness management.

In this article, we have proposed a trustworthy object classification framework, known as \emph{Trust--SIoT} by employing the concept of social trust theory to address the above-mentioned weaknesses. The main contributions of the proposed paper are as follows:

\begin{itemize}
    \item A trustworthy object classification framework (\emph{Trust--SIoT}) is envisaged by employing the social characteristics of objects in terms of direct trust metrics, reliability and benevolence, credible recommendations, and the degree of relationships. 
    \item We construct the SIoT knowledge graph to record five dynamic social relationships of SIoT objects in order to measure the degree of relationships. The learned graph embedding vectors are then used to estimate the object's social similarity using a knowledge graph embedding (KGE) model.
    \item 
    %Then, 
    Using complex non-linear correlations to map the input (i.e., trust measures) and output, we have proposed an artificial neural network-based heuristic for trustworthy object classification. 
    \item Finally, the performance of the proposed framework is evaluated by a number of evaluation metrics, i.e., F1-score, MAE, and MSE on two real-world datasets against the existing state-of-the-art.
\end{itemize}

The remainder of the paper is organized as follows: Section \ref{sec:back_rel_work} presents the discussion on SIoT, generalized trust background and the existing trust management models for SIoT and online social networks. In Section \ref{sec:prob_setup}, we have defined the problem
%setup, 
and introduced the used notations. Section \ref{sec:trust_quqn} delineates the proposed trust quantification framework for SIoT, and Section \ref{sec:exp_res} reports the experimental setup and experimental results from the performance evaluation 
%the performance 
of the proposed model. Finally, Section \ref{sec:conc} gives the concluding remarks and future research directions.

\section{Background and Related Work}
\label{sec:back_rel_work}

\subsection{Social Internet of Things (SIoT)}

The SIoT is the convergence of two domains: IoT and social networks. As a result, the paradigm of SIoT is formed wherein object establish the social links to create the social network of objects. In SIoT, an object is not only capable of measuring the surrounding ambience but can also establish the social relationships with other objects in the network on its own. These relationships enable an object to exchange the information with other objects (i.e., friends) in a trustworthy manner, and thus, provide the confidence of information sources as objects' tends to trust the friends having common characteristics, similar to humans (i.e., \emph{homophily} theory \cite{10.1145/1386790.1386838}). In the SIoT, an object can establish a variety of relationships, including but not limited to ownership object relationships (OOR), which are formed between objects belonging to the same owner (e.g., a mobile phone and a laptop), parental object relationships (POR), which are formed between objects manufactured by the same manufacturer and built at the same time (i.e., same batch), co-location and co-work object relationships (CLOR and CWOR), where CLOR represents the relationship established between the objects providing the services at the similar locations, whereas CWOR is established between the object providing the composite services, and social object relationships (SOR), 
%SOR 
which is established when two objects come into contact with each other because of their owners (e.g., mobile devices of two friends), and 
%it 
is dependent on the frequency of meeting between the owners's meeting.

\subsection{Trust in SIoT}

Trust is generally regarded as an integral constituent in any human social relationship. It is measured in terms of the degree of confidence pertinent to an individual's or any entity's likelihood to behave in an anticipated manner. The notion of trust varies across disciplines, i.e., trust has been studied by researchers in the domains of human sciences, i.e., sociology and psychology \cite{doi:10.1177/1368431003006002002} \cite{KRUEGER201992}, and in economics and computer science \cite{Backhouse2009} \cite{10.1145/642611.642634}. Each of these disciplines regard trust in a different perspective, and as such, their narrative could not be directly applied to SIoT. In SIoT, trust is referred to as the assessment of trustee's actions (or the assessment of data it provides) in view of trustor's expectations within a particular timespan. Furthermore, the assessment can be a probability, a value or a label, and be termed as trust esteem. In recent years, a plethora of trustworthiness management models are proposed, some of the latest literature is discussed in the subsection below. 

\subsection{Trustworthiness Models}

There exists a limited relevant work on trust management that employ prediction-based techniques (e.g., machine learning, deep learning), particularly, classifying the misbehaving objects who attack the reliable objects in the SIoT network \cite{9305298} \cite{9328463} \cite{9148767} \cite{sagar1}. Most recently, Marche \etal \cite{9305298} proposed a decentralized trust management model to detect the malicious objects that undertake various types of trust-related attacks. The suggested model, in general, employs machine learning heuristic to train the trust model on three trust scores: goodness, usefulness and perseverance. The suggested model's performance is compared to existing work using various trust-related attacks, and it is discovered that the proposed model performs well when using a combination of mix attacks, but decreases when using individual attacks. A matrix factorization-based approach is delineated by Aalibagi in \cite{9328463} wherein the SIoT network is portrayed as the bipartite graph of trustors and trustees providing and requesting different services. The model has exploited the Hellinger distance in order to propose a similarity and centrality-based social trust model. Finally, the matrix factorization approach is designed to extract the latent features of SIoT objects to classify the trustworthy objects, and to address the challenges of data sparsity and cold start problem. Sagar \etal \cite{sagar1} presented a social trust computational model by employing various types of similarities in terms of community-of-interest, friendship, and co-work similarity, and cooperativeness via social cooperation between SIoT objects. The trust model is then trained using a machine learning method (i.e., random forest) to classify the objects as trustworthy or 
%as 
untrustworthy.

Trust is also an important factor in social theory (i.e., social network) \cite{9142365}. A number of trust model for online social networks that may be extended for SIoT are presented in the literature, and they provide the extensive insight into the complex nature of social trust in terms of various trust properties, e.g., context-dependency, dynamic nature, uncertainty, transitivity, etc. The current literature employ a variety of approaches (neural network, walk-based, matrix factorization, etc.) to evaluate the trustworthiness of a node in online social network. Lin \etal \cite{9155370} designed a trust model based on node's  popularity and engagement as a trust features. Then, a graph convolutions neural network-based trust model is proposed to train the trust model in order to classify the trustworthy objects. The performance evaluation has shown the promising result when compared to the current existing state-of-the-art. Furthermore, a variety of walk-based approaches are delineated to assess the trust of a node by considering the trust propagation along the path from the trustor to the trustee \cite{8057106} \cite{6848107} to evaluate the trust score of a trustee. Furthermore, these techniques utilize a range of trust propagation (i.e., breadth first search and multi-path algorithms) method to traverse through social network graph. Finally, the trust estimation is computed as the aggregation of trust propagation from multiple paths.

Furthermore, it is imperative to study the current emerging techniques, (e.g., knowledge graph embedding (KGE)) that are currently the mainstream approach for knowledge graph representation in term of entity and relationships \cite{KGSurvey,2017Survey} \cite{RotL} \cite{TransE}. As SIoT network consists of social relationships between the SIoT objects, and these relationships are important part of trust evaluation process, thus, KGE can be utilized to represent these social relationships as the learning vector in order to quantify the social strength between the objects. As of now, only a few studies have explored the concept of KGE in SIoT for similarity computation \cite{doi:10.1177/15501477211009047} \cite{8936977}. Chen \etal in \cite{8936977}  \cite{doi:10.1177/15501477211009047} constructed the time-aware SIoT knowledge graph to estimate the social similarity of heterogeneous SIoT relationships by embedding the SIoT relationships in low-dimensional vector. Then, these low-dimensional vectors are used to determine the social similarity of the objects in order to provide the service/object recommendations.

In general, SIoT can benefit from the state-of-the-art trust evaluation methodologies and the trust properties utilized in online social networks, and in this paper, we have employed a number of such properties (i.e., reliability and benevolence) that have been explored in social networks. Furthermore, it is also imperative 
%to 
to explore the emerging techniques to effectively utilize SIoT paradigm's social aspects. Thus, we have exploited the low computation knowledge graph embedding (KGE) model to measure the social similarity of SIoT objects in terms of degree of relationships in the proposed framework. 

\section{Problem Setup}
\label{sec:prob_setup}

\begin{figure}[!tb]
    \centering
    \includegraphics[scale=0.4]{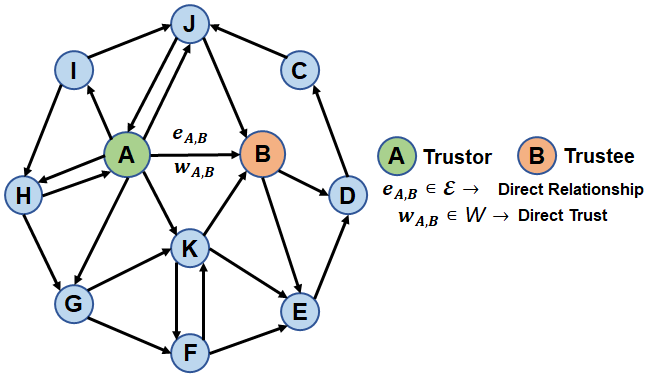}
    \caption{SIoT network as a directed graph $\mathcal{G} = (\mathcal{V},\mathcal{E},\mathcal{W})$}
    \label{fig:siot_network}
\end{figure}

\begin{figure*}[h]
    \centering
    \includegraphics[scale=0.5]{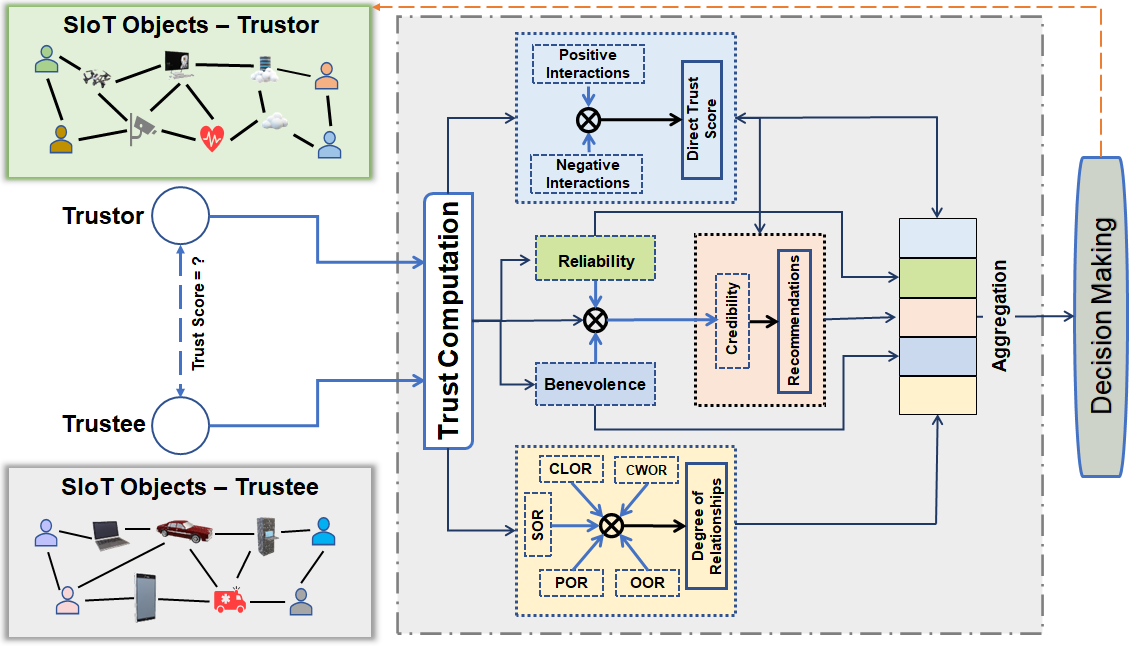}
    \caption{Illustration of \emph{Trust--SIoT} framework ($\otimes$ represents concatenation)}
    \label{fig:trust_framework}
\end{figure*}

In this paper, we consider the trust quantification of a social object in a SIoT network which is modeled in the form of a directed graph as $\mathcal{G} = (\mathcal{V},\mathcal{E},\mathcal{W})$ (Figure \ref{fig:siot_network}), wherein $s_i,s_j \in \mathcal{V}$ represents the social objects, the edge $e_{s_i,s_j} \in \mathcal{E}$ manifests the direct relationships, and $w_{s_i,s_j} \in \mathcal{W}$ represents the evaluation function in terms of trust relationship of social object (trustor-trustee) pair $(s_i,s_j)$. Let $\mathcal{W}^t = \{\mathcal{T}(s_i,s_j), w_{s_i,s_j}|e_{s_i,s_j} \in \mathcal{E}\}$ denotes the set of unobserved trust between any trustor-trustee pair to be quantified at any given time $t$. In order to quantify the trustworthiness of a social object, $object-object$ observations in terms of $successful$ and $unsuccessful$ interactions and social information of social objects via fundamental SIoT relationships are employed as our primary data source. The nomenclature is provided in Tabel \ref{tab:notation} for reference.

\renewcommand{\arraystretch}{1.5}
\newcolumntype{C}[1]{>{\centering\arraybackslash}p{#1}}
\newcommand{\centered}[1]{\begin{tabular}{l} #1 \end{tabular}}
\begin{table}[h]
\scriptsize
  \begin{center}
    \caption{Summary of notations}
    \label{tab:notation}
    \begin{tabular}{|>{\centering\arraybackslash}m{1.5cm} |>{\centering\arraybackslash}m{5.5cm}|}
     \hline
      \small\textbf{Notation} & \small\textbf{Description} \\ 
      \hline 
        $\mathcal{P}$ & Positive (Successful) Interactions \\ \hline
        $\mathcal{N}$ & Negative (Unsuccessful) Interactions \\ \hline
        $\mathcal{B}$ & Benevolence \\ \hline
        $\mathcal{R}$ & Reliability \\  \hline
        $\mathcal{CR}$ & Credibility \\  \hline
        $\mathcal{K}$ & Number of credible objects \\ \hline
        $\Phi$ & Trust Factor \\ \hline
        $\lambda$ & Rate of Trust Factor \\ \hline
        $f$ & KGE scoring function \\ \hline
        $\mathcal{L}$ & Cross Entropy Loss \\ \hline
        $\mathcal{C}-\mathcal{D}$o$\mathcal{R}$ & Context-aware Degree of Relationships \\ \hline
        $\mathcal{DTM}$ & Direct trust of trustor-trustee pair \\ \hline
        $\mathcal{RTM}$ & Recommendations for trustee \\ \hline
        $\mathbb{T}$ & Final Trust Score
          \\ \hline
    \end{tabular}
  \end{center}
  \vspace{-3mm}
\end{table}

Let $\mathcal{S} = \{s_1,s_2,...,s_n\}$ represents the set of social objects. Given a set of $object-object$ interactions and their social profiles in terms of a variety of relationships between them, the target problem of this research paper is to formulate the trust quantification by employing a number of trust metrics, i.e., Direct Trust Metric ($\mathcal{DTM}$), Reliability $\mathcal{R}$, Benevolence $\mathcal{B}$, Recommendations as a Trust Metric ($\mathcal{RTM}$), and Context-based Degree of Relationships ($\mathcal{C}-\mathcal{D}o\mathcal{R}$) in the form of a quintuple $<\mathcal{DTM},\mathcal{R}, \mathcal{B}, \mathcal{RTM},\mathcal{C}-\mathcal{D}o\mathcal{R}>$. 

Direct Trust Metric is measured when two object establish a direct relationship with each other via direct interaction between them, and as a result, a direct link $e_{s_i,s_j}$ is established in a graph with a weight $\mathcal{DTM}^t (s_i,s_j)$ refers to the direct of trust of $s_i$ towards $s_j$ at any time interval $t$. Recommendations as a Trust Metric is measured when two social object can only establish the relationship with each other by the word of mouth via objects' neighbours (i.e., friends) or via global reputation. Reliability is the fairness of object in providing recommendation to other objects and in contrast, benevolence represents on how other objects believe the object providing recommendations. Context-based Degree of Relationships ($\mathcal{C}-\mathcal{D}o\mathcal{R}$) refers to the context-based social strength between two social objects interacting with each other. It represents the confidence of information sources as objects tend to have more confidence on the objects similar to themselves in respect to a variety of qualities and characteristics. In general, the pairwise trust between social objects is quantified as:
\begin{equation}
\mathbb{T} (s_i,s_j) = \{\mathcal{T} (s_i,s_j), \forall \ s_i,s_j \in \mathcal{V},\mathcal{T} \in [0,1]\}
\end{equation}
\begin{equation}
    \mathcal{T} (s_i,s_j) \leftarrow \mathcal{DTM} \oplus \mathcal{R} \oplus \mathcal{B} \oplus \mathcal{RTM} \oplus \mathcal{C}-\mathcal{D}o\mathcal{R}
\end{equation}
Here $\oplus$ is the aggregation operator.

\section{Trust Quantification Model for SIoT}
\label{sec:trust_quqn}

This study proposes the \emph{Trust--SIoT} framework for the quantification of trust vis-\'a-vis any two social objects. The proposed framework consists of five trust metrics, direct trust metric, reliability and benevolence, recommendations as a trust metric, and the degree of relationships. The high-level overview of the proposed framework is depicted in Figure \ref{fig:trust_framework}.

For the trust quantification, at first, the proposed framework computes the direct trust metric by employing positive and negative interactions. The reliability and benevolence are then measured to observe how the trustee is known amongst the neighbours in the network. Subsequently, the reliability and benevolence are integrated to measure the credibility of neighbours for recommendations, and the neighbours with a credibility score above the threshold being considered to provide the recommendations for the trustee. The social resemblance of both the trustor and trustee in terms of their social relationships is measured by employing knowledge graph embedding to learn the embedding vector in order to quantify the degree of relationships (i.e., the social similarity of the trustor-trustee pair). Finally, an artificial neural network-based (ANN-based) model is trained for decision-making to identify the trustee's level of trustworthiness (trustworthy, neutral, or untrustworthy) of the trustee. The following are the details of 
%each 
the trust metrics:

\subsection{Direct Trust Metric}

Direct trust metric ($\mathcal{DTM}$) represents a direct trust score for a trustor-trustee pair. The proposed model adopts the Bayesian inference model \cite{Ismail2002TheBR} with beta probability density function to quantify the direct trust of a social object (Eq. \ref{eq:direct_trust}). The direct trust of a social object $s_x$ (trustor) towards another object $s_y$ (trustee) is defined as:
\begin{figure}[!t]
    \centering
    \includegraphics[scale=0.5]{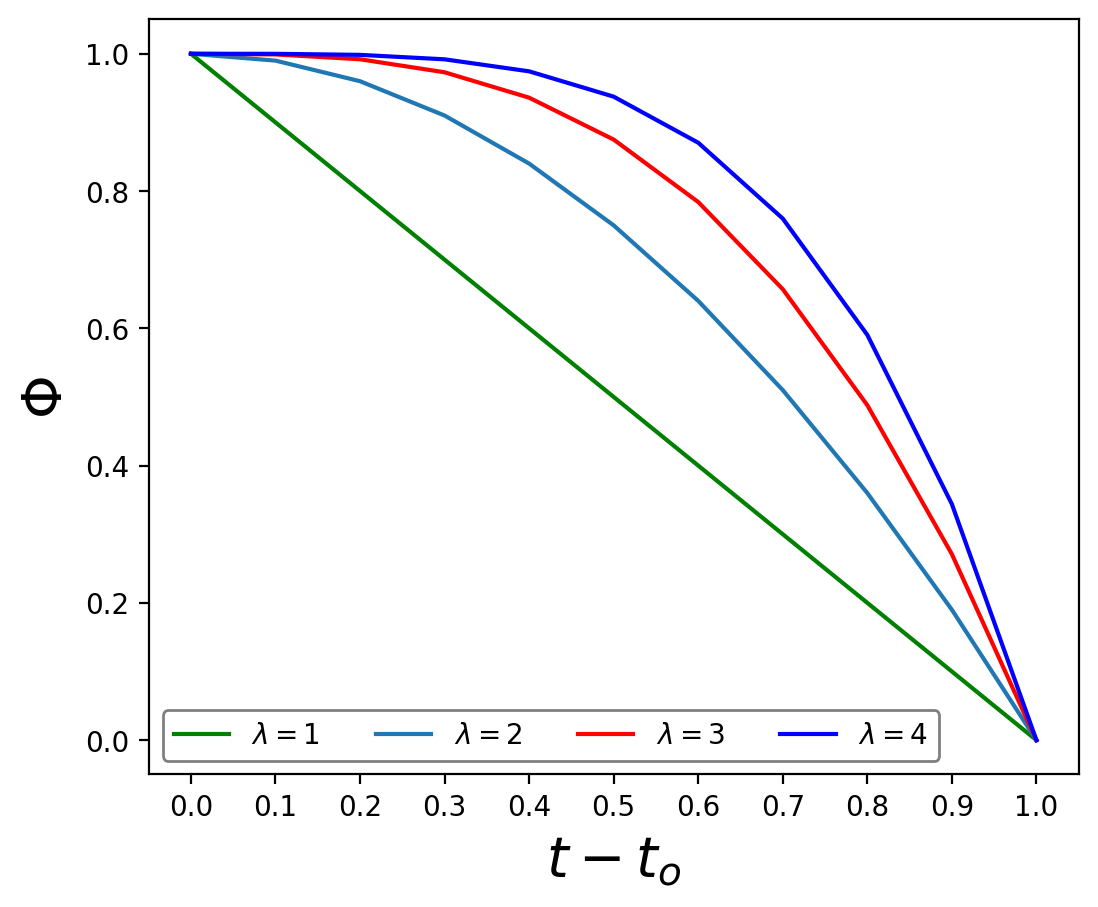}
    \caption{Decay in trust of a SIoT Object vis-\'a-vis time}
    \label{fig:past-Exp}
\end{figure}
\begin{equation}
% \begin{split}
\mathcal{DTM}^t (s_i,s_j) = \frac{\mathcal{P}^t_{s_i,s_j}+1}{\mathcal{P}^t_{s_i,s_j}+\mathcal{N}^t_{s_i,s_j}+2}
\label{eq:direct_trust}
\end{equation}
or, it can be simplified as:
\begin{IEEEeqnarray}{rCl}
\mathcal{DTM}^t (s_i,s_j) = \begin{cases}
\frac{\mathcal{P}^t_{s_i,s_j}}{\mathcal{P}^t_{s_i,s_j}+\mathcal{N}^t_{s_i,s_j}} & \forall \ \mathcal{P}^t_{s_i,s_j},\mathcal{N}^t_{s_i,s_j} > 0 \\
0.5  & otherwise
\end{cases}
% \IEEEyessubnumber
\end{IEEEeqnarray}
wherein, $\mathcal{P}^t_{s_i,s_j}$ denotes the number of the positive (successful) interaction carried out between trustor($s_i$)-trustee($s_j$) pair. Similarly, $\mathcal{N}^t_{s_i,s_j}$ denotes the count of negative (unsuccessful) interactions from the beginning to the current $t$.  

\begin{figure}[t]
    \centering
    \includegraphics[scale=0.55]{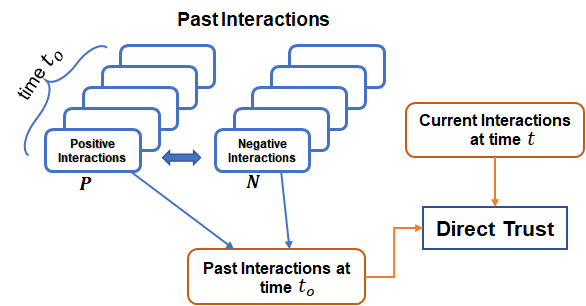}
    \caption{Direct trust computation}
    \label{fig:direct_trust}
\end{figure}

Past experience (interactions) from the beginning may not accurately reflect the behaviour of an object owing to the fact that an object's behaviour may change over time. Therefore, it is indispensable to model past experience by taking into account the feasible time interval that can provide the appropriate reputation of an object. The proposed model considers a trust factor $\Phi$ that can be tweaked to include the recent reputation of an object and can be employed in conjunction with interactions (i.e., $\mathcal{P}$ and $\mathcal{N}$). Eq. (\ref{eq:post}) and Eq. (\ref{eq:neg}) demonstrate the quantification of positive and negative interactions at any time $t$ with trust factor $\Phi$.
\begin{equation}
\begin{split}
\mathcal{P}_{s_i,s_j} & =  \Phi \times \mathcal{P}_{s_i,s_j}^{t-t_o} + \mathcal{P}_{s_i,s_j}^t \\
 & =  \left[1-(t-t_o)^\lambda \right] \times \mathcal{P}_{s_i,s_j}^{(t-t_o)} + \mathcal{P}_{s_i,s_j}^t
\end{split}
    \label{eq:post}
\end{equation}
\begin{equation}
\begin{split}
\mathcal{N}_{s_i,s_j} & = \Phi \times \mathcal{N}_{s_i,s_j}^{t-t_o} + \mathcal{N}_{s_i,s_j}^t \\
 & = \left[1-(t-t_o)^\lambda \right] \times \mathcal{N}_{s_i,s_j}^{(t-t_o)} + \mathcal{N}_{s_i,s_j}^t
\end{split}
    \label{eq:neg}
\end{equation}
Here, $(t-t_o)$ denotes the time interval that is considered to include the past interactions and $\lambda$ is the rate of trust factor. It can be observed from Figure \ref{fig:past-Exp} that change in the rate of trust factor has the influence on the gradual increase or decrease of trust factor. Similarly, when the value to $(t-t_o)$ is closed to zero, it means the model is only considering the latest reputation in terms of interactions of a social object and thus, the trust factor factor is higher. In general, higher trust factor signifies that the model is considering only the recent interactions, and this factor can be tuned by setting up different $t_o$ and $\lambda$. In general, the direct trust is the integration of current and past interactions (see Figure \ref{fig:direct_trust}).  

\subsection{Reliability and Benevolence}

In this section, we discuss two measure, \emph{reliability} and \emph{benevolence} by using the concept introduced in \cite{7837846}. Later, these measures will be used to compute the credibility of an object. 

\emph{Reliability} refers to the fairness of the object in providing a fair recommendation to the other objects in the network, an object providing a dishonest recommendation to a benign object must be considered less reliable. \emph{Benevolence}, on the other hand, represents the measure of an object on how other objects in the network trust or believe the object providing the recommendations. A benevolent object is believed to the trustworthy and intended to get a high rating from reliable objects. In general, both the reliability and the benevolence score depend on each other and can be computed recursively. The reliability and benevolence of an object $S$ can be quantified as: 
\begin{equation}
       \mathcal{B}^t(S) = \frac{1}{|InDeg(S)|} \sum\limits_{i \in InDeg(S)}{} \mathcal{R}^{t-1}(i) \times \mathcal{DTM}^t (i,S)
       \label{eq:benevolence}
\end{equation}
\begin{equation}
    \mathcal{R}^t(S) =  1 - \frac{1}{2|OutDeg(S)|} \sum\limits_{i \in OutDeg(S)}{} |\mathcal{DTM}^t (S,i) - \mathcal{B}^{t}(i)|
    \label{eq:reliability}
\end{equation}
Here, $InDeg(S)$ and $OutDeg(S)$ denote the inward and outward edge of an object $S$, $\forall \ edges \in \mathcal{E}$.

It can be seen from Eq. (\ref{eq:benevolence}), measuring the benevolence of an object requires the reliability of the objects facilitating the rating as well as the direct trust between objects connected via incoming edges. The final score is the average of all the incoming edges. Similarly, the reliability score (Eq. (\ref{eq:reliability})) is computed using the difference between the benevolence score of the objects and their respective direct trust score, and averaging all such differences to get the final reliability score. Furthermore, the \emph{reliability} and \emph{benevolence} are utilized to compute the credibility score of an object, and is defined as the subjective assessment of the perceived quality of an object being trusted and accepted \cite{wells2021assessing}. Finally, the credibility of an object is quantified as the product of reliability and benevolence (Eq. (\ref{eq:cred})) , i.e., a credible object must be both reliable and benevolent.
\begin{equation}
\label{eq:cred}
\mathcal{CR}^t(\mathcal{S}) =  \mathcal{R}^t(\mathcal{S}) \times \mathcal{B}^t(\mathcal{S})
\end{equation}

Algorithm \ref{alg:credibility} shows the steps involved for the computation of the credibility. The initial score of reliability and benevolence is set to $0.5$ (line 3). Line 7 to line 9 computes the benevolence of an object using the reliability score from the previous iterations. Likewise, line 10 to line 13 computes the reliability score of an object using the current benevolence score. The computation keeps going on until the convergence (i.e., the difference between both the score in two consecutive iterations is less than the error $\epsilon$) (line 6), and finally, the credibility is computed as the product of reliability and benevolence at line 14. For the demonstration of the credibility algorithm, consider a sample SIoT network (Figure 5) that has five objects. The set of objects $\{A,C,E\}$ are the credible objects and objects $\{B,C\}$ are not the credible objects. As can be seen from the reliability, benevolence, and credibility score in Table \ref{table:cred_score}, it is observed that credible objects need to be both reliable and benevolent. Though object $B$ and object $D$ have a high reliability score, the benevolence score for both these objects is very low which leads to a low credibility score.   

\begin{algorithm}[t]
\caption{$\mathcal{R},\mathcal{B},$ and $\mathcal{CR}$ of a SIoT Object $\mathcal{S}$ ($\forall \ \mathcal{S} \in \mathcal{V}$)}
\begin{algorithmic}[1]
  \STATE \textbf{Input}: SIoT Network as a Directed Graph $\mathcal{G} = (\mathcal{V},\mathcal{E},\mathcal{DTM)}$, $\mathcal{N}(\mathcal{S})$: Neighbours of an object $\mathcal{S}$, $InDeg(\mathcal{S})$ and $OutDeg(\mathcal{S})$ as in-degree and out-degree of an object $\mathcal{S}$ respectively.
  \STATE \textbf{Output}: $Reliability \ (\mathcal{R})$, $Benevolence \  (\mathcal{B})$ and $Credibility \ (\mathcal{CR})$
% \STATE Let $\mathcal{N}(S)$ represent neighbours of an object $s_i$, $InDeg(S)$ and $OutDeg(S)$ denote in-degree and out-degree of an object $S$ respectively
 \STATE Let $\mathcal{R}^0(\mathcal{S}) = 0.5$ and $\mathcal{B}^0(\mathcal{S}) = 0.5$
   \STATE $t=1$
   \STATE $k \in \{ N(\mathcal{S}) , \forall \ N(\mathcal{S}) \in \mathcal{V}\}$  
  \WHILE{$|\mathcal{R}^t(k)-\mathcal{R}^{t-1}(k)| > \epsilon$ $||$ $|\mathcal{B}^t(k)-\mathcal{B}^{t-1}(k)| > \epsilon$}
  \FOR{$s_{ik} \in InDeg(k)$, $\forall s_{ik} \in \mathcal{V}$}
  \STATE $\mathcal{B}(s_{ik})^{t} = \frac{1}{|s_{ik}|} \sum_{s_{ik}} \mathcal{R}(k)^{t-1} * \mathcal{DTM}^t (s_{ik},k)$
  \ENDFOR
  \FOR{$s_{ok} \in OutDeg(k)$, $\forall s_{ok} \in \mathcal{V}$}
  \STATE  $\mathcal{R}(k)^{t} =  1-\frac{1}{2|s_{ok}|} \sum_{s_{ok}} \mathcal{DTM}^t (k,s_{ok}) - \mathcal{B}(s_{ok})^{t-1}$
  \ENDFOR
  \ENDWHILE
  \STATE $\mathcal{CR}^{t} (k) = \mathcal{R}(k)^{t} * \mathcal{B}(k)^{t}$
  \STATE \textbf{return} $\mathcal{CR}^{t}(k)$
\end{algorithmic}
\label{alg:credibility}
\end{algorithm}

\begin{table}[ht]
\begin{minipage}[b]{0.5\linewidth}
\centering
\includegraphics[scale=0.6]{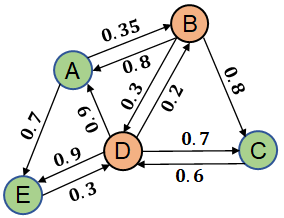}
\captionof{figure}{Sample SIoT Network}
\label{fig:sample_siot}
\end{minipage}\hfill
\begin{minipage}[b]{0.5\linewidth}
\centering
\scriptsize
\begin{tabular}{ c | c | c | c }
    \hline
    $S$ & $\mathcal{R}(S)$ & $\mathcal{B}(S)$ & $Cr(S)$\\ \hline \hline
    A & 0.96 & 0.82 & 0.79 \\ \hline
    B & 0.97 & 0.26 & 0.25 \\ \hline
    C & 0.88 & 0.72 & 0.63 \\ \hline
    D & 0.96 & 0.37 & 0.35 \\ \hline
    E & 0.96 & 0.76 & 0.73 \\ \hline

   \end{tabular}
    \caption{Credibility of $\mathcal{S}$}
    \label{table:cred_score}
\end{minipage}
% \captionsetup{labelformat=empty}
% \caption{Fig. \ref{fig:sample_siot}: A sample of SIoT network for credibility evaluation, Table \ref{table:cred_score} Reliability, benevolence and credibility score.}
\end{table}

\subsection{Recommendation as a Trust Metric} 

Recommendation as a trust metric ($\mathcal{RTM}$) is indispensable when the direct information is insufficient to evaluate the trustworthiness of an object. Therefore, a reliable recommendation trust is essential to provide the requisite trust recommendations at any time $t$. In distributed SIoT network, the recommendations of an object is provided by the trustor's neighbours (i.e., local recommendations) having a direct relationship with the trustee or via the global reputation of a trustee. Nevertheless, to mitigate the malicious behaviour of objects providing dishonest recommendations (i.e., good and bad mouthing attack), we have proposed reliability ($\mathcal{R}$) and benevolence ($\mathcal{B}$) based credibility score ($\mathcal{CR}$) computation method (\ref{sec:cred_comp}) to select the top credible objects before relying on the recommendations provided by them.  

% \label{sec:cred_comp}
An object (i.e., trustor) $s_i$ compute the credibility of all the $k \in \{k, \forall k \in \mathcal{V}\}$ recommenders providing the recommendation trust for another object (i.e., trustee) $s_j$. The credibility ($\mathcal{CR}^t(S)$) of an object $S$ at any time $t$ is computed as the product of reliability and benevolence of that object (Eq. (\ref{eq:cred}))

As a whole, the top $\mathcal{K} = \{\mathcal{K}, \forall \ \mathcal{K} \in \mathcal{V}\}$ credible objects having credibility above the pre-defined threshold $(Th)$ (Eq. (\ref{eq:credible_x})) are selected to provide the recommendation for a trustee.
\begin{equation}
\mathcal{K} = \begin{cases}
\mathcal{K} & \mathrm{if} \ \mathcal{CR}^t (\mathcal{K}) \geq Th \\
\mathcal{K} \in \{\} & otherwise
\end{cases}
\label{eq:credible_x}
\end{equation}

Finally, the recommendation trust ($\mathcal{RTM}^t(s_i,s_j)$) can be computed as sum of the product of the direct trust of credible objects for the trustee ($\mathcal{DTM}^t(\mathcal{K},s_j)$) and the direct trust of trustor for the credible objects ($\mathcal{DTM}^t(s_i,\mathcal{K})$), and the final score is the average of all recommendations from $\mathcal{K}$.   
\begin{equation}
 \mathcal{RTM}^t (s_i,s_j)  = \frac{1}{|\mathcal{K}|}  \sum\limits_{i=1}^{|\mathcal{K}|} \mathcal{DTM}^t(s_i,i) \times \mathcal{DTM}^t(i,s_j)  
\end{equation}

Furthermore, if the local recommendation is not present for a trustee at a specific time $t$, then the global recommendation (the overall reputation of a trustee in the SIoT network ($\mathcal{G}$)) is considered to compute the recommendation ($G-\mathcal{RTM}$). The global recommendation is computed in Eq. (\ref{eq:rec_gloabl}).
\begin{equation}
     \mathcal{RTM}^t (s_i,s_j)  = \frac{1}{|\mathcal{U}|}  \sum\limits_{i=1}^{|\mathcal{U}|} \mathcal{DTM}^t(i,s_j)
     \label{eq:rec_gloabl}
\end{equation}
Here $\mathcal{U} = \{\mathcal{U}, \forall \ \mathcal{U} \in \mathcal{V}\}$ denotes the number of objects having at least one transaction with the trustee in the entire network.

\subsection{Knowledge Graph Embedding-based Degree of Relationships}

Context-aware Degree of Relationship ($\mathcal{C}-\mathcal{D}$o$\mathcal{R}$) is one of fundamental components of SIoT as objects maintain a number of relationships to interact with each other just as human beings do \cite{MS201932}. These relationships are important in modeling the trustworthiness of an object in SIoT network as object tends to trust the objects having similar behaviour and characteristics \cite{ATZORI20123594}. For $\mathcal{C}$-$\mathcal{D}$o$\mathcal{R}$ quantification, the proposed model has considered following relationships from the public SIoT dataset \cite{marche2020exploit}:

\begin{itemize}
    \item[1)] \emph{CLOR}: If two or more objects provide the services at the similar location, then these objects form this type of location-based relationships.  
    \item[2)] \emph{POR}: this type of relationships is correlated with similar objects having same manufacturer and are built within same period of time. 
    \item[3)] \emph{OOR}: OOR represents the relationships between the objects of the same owners (i.e., devices of same owner). 
    \item[4)] \emph{SOR}: When two or more objects come into contact with each other, they form a social relationship similar to that of humans. This type of relationships are based on number of meetings, meeting duration, and interval between two consecutive meetings. In the mentioned dataset, this relationship is created between private mobile devices.
    \item[3)] \emph{SOR$_2$}: This relationship is the variant of \emph{SOR} and it is created to connect public and private mobile devices.  
\end{itemize}

Given the above five types of relationships, we propose a knowledge graph embedding (KGE) method for estimating social object similarity in terms of 
$\mathcal{C}$-$\mathcal{D}$o$\mathcal{R}$.
The KGE model is currently one of mainstream techniques for knowledge graph representation learning, which aims to represent each entity $e \in E$ (or relation $r \in R$) as a $d$-dimensional continuous vector, denoted as $\mathbf{e}$ (or $\mathbf{r}$). The entity embedding vectors outputted by the KGE model record the structure features of SIoT objects in the multi-relational graph and can be utilized to compute the object similarity.

\begin{figure}[!t]
    \centering
    \includegraphics[width=0.9\linewidth]{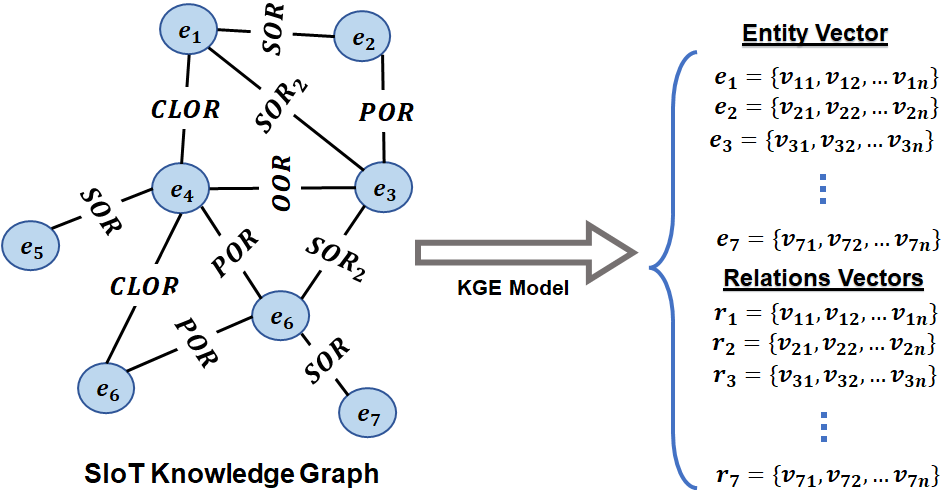}
    \caption{SIoT knowledge graph}
    \label{fig:siot_kge}
\end{figure}

Specifically, given the SIoT objects and their relational data according to the five relationships, we define the SIoT knowledge graph as a multi-relational graph $\mathcal{G}_k= (\mathcal{V},\mathcal{R})$ (Figure. \ref{fig:siot_kge}) that can be represented as a set of triples $T=\{(e_h,r,e_t)\}$, where head entity and tail entity $e_h,e_t \in \mathcal{V}$ are the object nodes and $r \in \mathcal{R}$ is one of the five relations between objects. 
$n_T$, $n_V$, $n_R$ represent the number of triples, entities and relations in the graph $\mathcal{G}_k$ respectively.

Considering the scalability of the proposed model in SIoT, we pay special attention to the training complexity and parameter amount of knowledge graph embeddings.
To this end, we review the recent low-dimensional KGE models \cite{GoogleAttH,MulDE,RotL} and select the RotL model\cite{RotL}, a lightweight Euclidean-based KGE model. 
Unlike previous low-dimensional KGE models requiring complicated calculations in hyperbolic space, this model can process different kinds of relations and obtain high prediction accuracy in low-dimensional Euclidean vector space. 
The core part of the KGE model is the scoring function, which calculates the triple score based on the embedding vectors of the triple items. A higher score means this triple is more likely to be true. 
Given a KG triple $(e_h, r, e_t)$, The scoring function of RotL is defined as:
\begin{align}
f(e_h, r, e_t) = -\varphi(\|Rot(\mathbf{r})\mathbf{e_h} \oplus_\alpha \mathbf{e_t}\|) + b_{e_h} + b_{e_t}
\end{align}
where $\varphi(x)=xe^x$ is a nonlinear activation function, $Rot(\cdot)$ is a Givens Rotation operation, the addition operation $\oplus_\alpha$ is defined as $\mathbf{x} \oplus_\alpha \mathbf{y} = \alpha(\mathbf{x}+\mathbf{y})/(1+\mathbf{x}\mathbf{y})$, $\alpha$ is a relation-specific trainable parameter and $b_e (e\in E)$ are trainable entity biases.

In order to train the embedding parameters in the RotL model, 
we exploit the negative sampling loss which is the commonly-used training strategy in the KGE field.
Given a triple $t = (e_h, r, e_t) \in T$, a random sampling strategy is first used to replace one entity in the triple and generate a set of negative samples $T'_t$:
\begin{align}
    T'_t = \{(e_h, r, e_t') \mid e_t' \in E \land e_t' \neq e_t\} \cup \\ \nonumber
    \{(e_h', r, e_t) \mid e_h' \in E \land e_h' \neq e_h\}
\end{align}
After that, the negative sampling loss forces the score of the correct triple higher than those of its negative samples, so as to learn effective vector representations for the KG. 
Given the training triple set $T_{tr}=\{t\}$, in which each triple $t$ has a set of negative triples $T'_t$, the loss function based on the binary classification cross entropy loss is as follows:
\begin{align}
\mathcal{L}_{KGE}=-\frac{1}{n_{T_{tr}}} \sum_{t=(e_h,r,e_t)\in T_{tr}} (log(\sigma(f(e_h,r,e_t)))  \\ \nonumber
+ \sum_{(e_h',e_r',e_t')\in T'_t}log(1-\sigma(f(e_h',e_r',e_t')))),
\end{align}
where $\sigma(\cdot)$ denotes the Sigmoid function. Following the original settings of the RotL model, we utilize minimize the loss by exploiting the Adam optimizer \cite{ADAM}.

After training the RotL model using the SIoT KG triples, we can obtain the low-dimensional embedding vector $\mathcal{O}_v(S)$ of each object ($\mathcal{S}$) in the SIoT network. Finally, we estimate the social similarity as the $\mathcal{C}-\mathcal{D}o\mathcal{R}$ vis-\'a-vis objects (Eq. (\ref{eq:cdor})).
\begin{equation}
\label{eq:cdor}
\mathcal{C}-\mathcal{D}o\mathcal{R}(s_i,s_j) = \frac{\mathcal{O}_v(s_i).\mathcal{O}_v(s_j)}{||\mathcal{O}_v(s_i)|| \ ||\mathcal{O}_v(s_i)||}
\end{equation}

\subsection{Final Trust Score}

% We use Neural Networks
% (NN) [13], a machine learning algorithm, to determine the
% model T . Our choice of Neural Networks is based on two
% reasons: (1) Neural Networks can detect complex non-linear
% relationships between input and output variables [14] (2) the
% significance of the trust attributes can easily be computed
% using the trained model
This section introduces the heuristic to aggregate all the quantified trust metrics to classify the level the trustworthiness of an object. Traditionally, the existing works utilize the weighted average mean of the trust metrics (Eq. (\ref{eq:wei_trust})) to obtain the level of trustworthiness. However, selecting the appropriate weights for each metric is still a challenge as there could be an infinite number of possibilities to select the optimal weight. 
\begin{equation}
\begin{aligned}
    \mathcal{T} (s_i,s_j) = {} & w_1*\mathcal{DTM} \oplus w_4*\mathcal{R} \oplus w_5*\mathcal{B} \oplus w_2*\mathcal{RTM} \oplus \\ 
    & w_3*\mathcal{C}-\mathcal{D}o\mathcal{R} 
\end{aligned}    
\label{eq:wei_trust}
\end{equation}
where $w_1$-$w_5$ represent the weight for each metric and the sum $w_1+w_2+w_3+w_4+w_5 = 1$ to get the trust score in the range $\{0,1\}$. $\oplus$ denotes the concatenation operation. Furthermore, the final trust is computed as:
\begin{equation}
\mathbb{T} (s_i,s_j) = \{\mathcal{T} (s_i,s_j), \forall \ s_i,s_j \in \mathcal{V},\mathcal{T} \in [0,1]\}
\end{equation}

\begin{figure}[!t]
    \centering
    \includegraphics[scale=0.37]{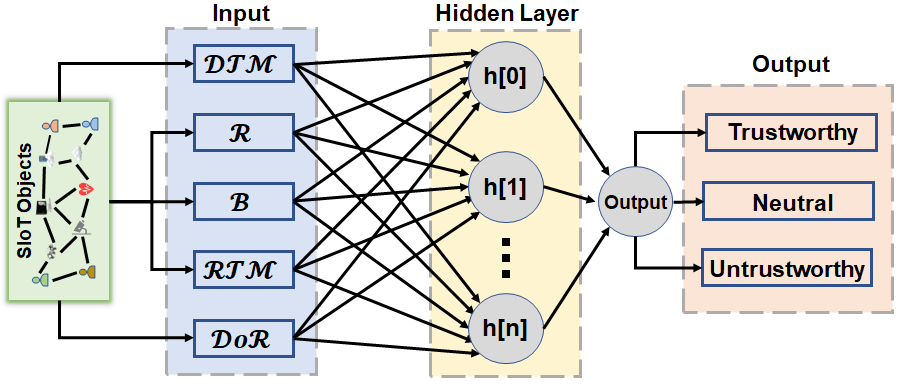}
    \caption{ANN-based trust training model}
    \label{fig:final_trust}
\end{figure}

To cope with the issue of selecting a weight, we have introduced an ANN-based heuristic to synthesize the proposed trust metrics by learning the weights for each metric. The high-level overview of how the trust metrics are employed to train the ANN model is depicted in Figure \ref{fig:final_trust}. There are a total of five trust metrics are employed to feed in the input layer, these 5-metrics act as 5 neurons. The neuron in the output layer represents the level of trustworthiness for each object in the SIoT network $\mathcal{G}$. In the proposed framework, we have classified the objects into three levels: trustworthy, neutral, and untrustworthy.   

Algorithm \ref{alg:ANN_train} summarizes the steps of training the ANN-based trust model. The training model takes trust metrics ($\mathcal{TM}$), data size ($N$), and cost threshold ($c$) as the input and returns the trained model ($\mathbb{T}$). Lines from 3 to 5 compute the contribution of each trust metric. The learning weights are computed using Adam's optimizer with the default cost threshold and subsequently, the difference between predicted trust and ground truth is used as the new cost value in each iteration from lines from 7 to 10. Finally, when the condition $cost>c$ satisfies, the loop terminates and returns the trained model $\mathbb{T}$.    

\begin{algorithm}[t]
\caption{Training ANN-based Trust Model}
\begin{algorithmic}[1]
  \STATE \textbf{Input}: Trust Metrics ($\mathcal{TM}$) $\{\mathcal{DTM},\mathcal{RTM},\mathcal{C}-\mathcal{D}o\mathcal{R},\mathcal{R},\mathcal{B}\}$, Data Size ($N$), Cost Function Threshold ($c$)
  \STATE \textbf{Output}: $\mathbb{T}$: Trained Trust Model
  \FOR{TM $\in \mathcal{TM}$}
  \STATE Compute the contribution of each TM.
  \ENDFOR
  \STATE Let $cost=\infty$
  \WHILE{$cost>c$}
  \STATE Compute the ANN weights for each $TM \in \mathcal{TM}$ for $N$ data samples using Adam's optimizer.
  \STATE Update the cost value (i.e., $cost$) based on newly learned weights 
  \ENDWHILE
  \STATE \textbf{return} Trained Model $\mathbb{T}$
\end{algorithmic}
\label{alg:ANN_train}
\end{algorithm}

Conclusively, all the steps of trustworthiness management of the proposed framework are realized in Algorithm \ref{alg:final_trust_algorithm}. The algorithm of trustworthiness computation follows a number of steps, including the direct trust computation, reliability and benevolence, recommendation from credible objects, and the degree of relationships. Finally, the quantified metrics are employed in the ANN-based heuristic to train the trust model and to classify the trustworthy objects. 

\begin{algorithm}[!tb]
\caption{Trust Evaluation}
\begin{algorithmic}[1]
  \STATE \textbf{Input}: $\mathcal{P}_{s_i,s_j}, \mathcal{N}_{s_i,s_j}, \mathcal{R}, \mathcal{B}$
  \STATE \textbf{Output}: $\mathcal{DTM}, \mathcal{RTM} \ and \ \mathcal{C}-\mathcal{D}o\mathcal{R} \ \mathcal{T}_{Final}$
  \\ \textbf{Direct Trust Computation ($\mathcal{DTM}$)}
  \STATE Compute Positive Interactions: $\mathcal{P}_{s_i,s_j} = [1-(t-t_0)^{\lambda}] * \mathcal{P}_{i_s,s_j}^{t-t_o} + \mathcal{P}_{s_i,s_j}^t$
  \STATE Compute Negative Interactions: $\mathcal{N}_{s_i,s_j} = [1-(t-t_0)^{\lambda}] * \mathcal{N}_{s_i,s_j}^{t-t_o} + \mathcal{N}_{s_i,s_j}^t$
  
  \STATE $\mathcal{DTM}^t (s_i,s_j) = \frac{\mathcal{P}_{s_i,s_j}+1}{\mathcal{P}_{s_i,s_j}+\mathcal{N}_{s_i,s_j}+2}$ \\ 
  
  \textbf{Reliability ($\mathcal{R}$), Benevolence ($\mathcal{B}$), Credibility ($\mathcal{CR}$)}
 
   \STATE Compute $\mathcal{R}$ and $\mathcal{B}$ using Algorithm \ref{alg:credibility}
   \\
 \textbf{Recommendations Computation ($\mathcal{RTM}$)}
  
  \STATE Compute the Credibility of Trustors' Neighbours:
  \STATE $\mathcal{CR}(\mathcal{N}(s_i)) = \mathcal{R}(\mathcal{N}(s_i)) \times \mathcal{B}(\mathcal{N}(s_i)) $ from Algorithm \ref{alg:credibility}
  \STATE Select top $\mathcal{K}$ credible neighbours for providing $\mathcal{RTM}$
  \IF{$\mathcal{K}\notin\{\}$}
    \STATE $\mathcal{RTM}^t (s_i,s_j)  = \frac{1}{|\mathcal{K}|}  \sum\limits_{i=1}^{|\mathcal{K}|} \mathcal{DTM}^t(s_i,i) \times \mathcal{DTM}^t(i,s_j)$ // Local Recommendations
  \ELSE
  \STATE $\mathcal{RTM}^t (s_i,s_j)  = \frac{1}{|\mathcal{U}|}  \sum\limits_{i=1}^{|\mathcal{U}|} \mathcal{DTM}^t(i,s_j)$ \\ // Global Recommendations
  \ENDIF \\
  \textbf{Degree of Relationships ($\mathcal{C}-\mathcal{D}o\mathcal{R}$)} \\
  \STATE \textbf{Input:} Relationships vis-\'a-vis objects
  \STATE \textbf{Output:} Embedded vectors representing objects' relationships ($\mathcal{O}_v$)
  \STATE Similarity Computation as $\mathcal{C}-\mathcal{D}o\mathcal{R}$
  \STATE $\mathcal{C}-\mathcal{D}o\mathcal{R}(s_i,s_j) = \frac{\mathcal{O}_v(s_i).\mathcal{O}_v(s_j)}{||\mathcal{O}_v(s_i)|| \ ||\mathcal{O}_v(s_i)||}$ \\
  \textbf{Final Trust Classification ($\mathbb{T}$)} \\
  \STATE $\mathbb{T} = $\\ $w_1*\mathcal{DTM}+w_2*\mathcal{R}+w_3*\mathcal{B}+w_4*\mathcal{RTM}+w_5*\mathcal{C}-\mathcal{D}o\mathcal{R}$ where $w_1+w_2+w_3+w_4+w_5=1$
  \STATE Trust decision making via object classification (i.e., Trustworthy, Neutral and Untrustworthy) using Neural Network
\end{algorithmic}
\label{alg:final_trust_algorithm}
\end{algorithm}

\section{Experimental Setup and Results}
\label{sec:exp_res}

This section presents the extensive experimental evaluation of our proposed framework (\emph{Trust--SIoT}). We have implemented the algorithms using Python 3.6 on a MAC environment with an 8-core CPU and 16GB RAM. To further validate, we have compared the \emph{Trust--SIoT} framework with a well-known trust model named \emph{Guardian}\cite{9155370} from existing state-of-the-art. \emph{Guardian} combines two trust features to classify the trustworthy objects: \emph{popularity}, refers to the trustworthiness of an object observed by other objects, and \emph{engagement}, refers to the trustworthiness of others observed by objects' perceptive.    

\subsection{Experimental Settings}

\textbf{Datasets Description:} We conduct our experiments on two real-world datasets: \emph{Advogato}\footnote{http://www.trustlet.org/datasets/advogato/} and \emph{Bitcoin-Alpha\footnote{https://btc-alpha.com/en/} (BTC-Alpha)}. Advogato is an online social network of open source developers and it allows users to certify each other with different levels of trustworthiness. In particular, the different categories of trustworthiness are $\{Observer, Apprentices, Journeyer, Master\}$ with each having a numerical value of $\{0.1,0.5,0.7,1.0\}$. The BTC-Alpha dataset adopts the concept of Web-of-trust to provide safety and security by allowing users to rate each other they trust or distrust to maintain the record of the user's reputation. The site focuses on having an open market wherein users can make transactions with each other by using bitcoins, and respectively rate each other. The rating (or trust) in this dataset varies in the range of $\{0,1\}$. The statistical description of these datasets is presented in Table \ref{table:data_des}. 

\begin{table}[h]
    \centering
    \caption{Statistical Description of Datasets}
    \begin{tabular}{| c | c | c | c |}
    \hline
    $Datasets$ & $\# \ of \ Objects$ & $\# \ of \ Edges$ & $Avg. \ Degree$\\ \hline 
    \emph{Advogato}  & $6,541$ & $51,127$ & $19.2$ \\ \hline
    \emph{BTC-Alpha} & $3,775$ & $22,650$ & $12.79$ \\ \hline
   \end{tabular}
    \label{table:data_des}
\end{table}

\textbf{Data Preparation:} As the selected datasets do not contain the SIoT relationships information among the objects which is the fundamental component for any SIoT network, we have extracted the SIoT relationships information from \emph{social-dataset}\footnote{https://www.social-iot.org/}. The social-dataset contains the information of the real IoT objects deployed in Santandar city of Spain, and there are a total of $16,216$ IoT objects. Furthermore, this dataset includes the five different types of SIoT relationships (i.e., CLOR, POR, OOR, SOR, and SOR$_2$) between the IoT objects that connect them to form a SIoT network, and information about the service requested by these objects. We have selected the relationships information of a sub-network of this dataset to integrate it with the \emph{Advogato} and \emph{BTC-Alpha} by selecting the objects to enhance the probability of objects interacting with each other. The merging takes place in three steps: 1) conversion of SIoT dataset into SIoT graph, 2) selecting the sub-network of data to match the number of nodes for each of \emph{Advogato} and \emph{BTC-Alpha} datasets, and 3) merging the sub-network of SIoT data with both \emph{Advogato} and \emph{BTC-Alpha} dataset. Finally, we have mapped the labels of merged datasets into three classes $\{Trustworthy, Neutral, Untrustworthy\}$. 

For all the datasets, we have randomly chosen $80\%$ (i.e., training set) of the data for training, and the remaining $20\%$ of the data is used for testing (i.e., testing set). We further split the training set into 5 subsets for 5-fold cross-validation. Each time the training and validation phase chooses one of the 5 subsets for validation and the remaining for training and this step is repeated five times to pick the best average performance values. Furthermore, the testing set is used to evaluate the performance of the model on test data. We have implemented our framework by using grid search and 5-fold cross-validation, and with default hyperparameters (i.e., a single hidden layer, rectified linear unit function (relu) as the activation function, penalty (regularization term) of $0.0001$, and `adam' solver) to train our neural network.  

\textbf{Evaluation Metrics:} We employ three standard evaluation metrics (i.e., F1-score ($F1$), Mean Absolute Error ($MAE$), and Mean Squared Error ($MSE$)) to evaluate the prediction accuracy of our model in comparison with the existing state-of-the-art. In particular, we have used 
%michael; F!?? pls check
micro-averaged F1 score due to label imbalance in both the datasets. Note that higher value of $F1$ and lower values of $MAE$ and $MSE$ indicate better performance. 

\begin{itemize}
    \item \textbf{Mean Absolute Error:} 
    \begin{equation}
        MAE = \frac{1}{N_{test}} \sum_{s_i,s_j} | \mathbb{T}_{s_i,s_j} - \hat{\mathbb{T}}_{s_i,s_j}|
    \end{equation}
    \item \textbf{Mean Squared Error:}  
    \begin{equation}
        MSE = \frac{1}{N_{test}} \sum_{s_i,s_j} | \mathbb{T}_{s_i,s_j} - \hat{\mathbb{T}}_{s_i,s_j}|^2
    \end{equation}
    \item \textbf{F1-Score:}  
    \begin{equation}
        F1 = 2 \times \frac{Precision \times Recall}{Precision + Recall}
    \end{equation}
\end{itemize}
where $\mathbb{T}_{s_i,s_j}$ is the original trust score given by an object $s_i$ to another $s_j$ and $\hat{\mathbb{T}}_{s_i,s_j}$ is the predicted trust score of $s_i$ towards $s_j$. $N_{test}$ is the number of interactions in test data. 

\subsection{Results and Discussion}

Firstly, we have analyzed the \emph{effectiveness} of \emph{Trust--SIoT} on both the datasets. Table \ref{table:pred_acc_adv} and Table \ref{table:pred_acc_btc} portray the effectiveness of the both the approaches on \emph{Advogato} and \emph{BTC-Alpha} respectively. It is evident from Table \ref{table:pred_acc_adv} that \emph{Trust--SIoT} outperforms the \emph{Guardian} in terms of F1-score with with an improvement of $3.6\%$. Furthermore, \emph{Trust--SIoT} achieves the higher prediction accuracy in terms of MAE and MSE score of $0.174$ and $0.240$ in comparison to MAE and MSE score of $0.214$ and $0.3008$ for \emph{Guardian}. In general, \emph{Trust--SIoT} achieves higher F1-Score and $18.7\%$ and $20.2\%$ decrease in the MAE and MSE score in comparison to \emph{Guardian}. 

Similarly, we have evaluated the performance of our approach on another dataset (i.e., \emph{BTC-Alpha}) to verify that our approach does not rely on datasets. As can be seen from Table \ref{table:pred_acc_btc}, \emph{Trust--SIoT} performance slightly better than \emph{Guardian} by increasing the F1-score $0.5\%$ with a decrease in $1.9\%$ of MAE score and almost similar MSE score. In general, \emph{Trust--SIoT} is highly effective on both the datasets, and it successfully clarify that the trust metrics along with the SIoT relationships similarity obtained from KGE model are able to classify the trustworthy SIoT objects more efficiently.   

\begin{table}[h]
    \centering
    \caption{Prediction accuracy - \emph{Advogato}}
    \begin{tabular}{| c | c | c | c |}
    \hline
    $Model$ & $F1-Score$ & $MAE$ & $MSE$\\ \hline \hline
    \emph{Trust--SIoT}  & 0.859 & 0.174 & 0.240 \\ \hline
    \emph{Guardian} & 0.829 & 0.214 & 0.3008 \\ \hline \hline
    Improvement & \textbf{3.6\%} & \textbf{18.7\%} & \textbf{20.2\%} \\ \hline \hline
   \end{tabular}
    \label{table:pred_acc_adv}
\end{table}

\begin{table}[h]
    \centering
    \caption{Prediction accuracy - \emph{BTC Alpha}}
    \begin{tabular}{| c | c | c | c |}
    \hline
    $Model$ & $F1-Score$ & $MAE$ & $MSE$\\ \hline \hline
    \emph{Trust--SIoT}  & 0.785 & 0.211 & 0.220 \\ \hline
    \emph{Guardian} & 0.781 & 0.215 & 0.219 \\ \hline \hline
    Improvement & \textbf{0.5\%} & \textbf{1.9\%} & -0.4\% \\ \hline \hline
   \end{tabular}
    \label{table:pred_acc_btc}
\end{table}

\begin{figure}[!tb]
\centering
     \begin{subfigure}[b]{0.25\textwidth}
         \centering
         \includegraphics[width=4.5cm,height=3.5cm]{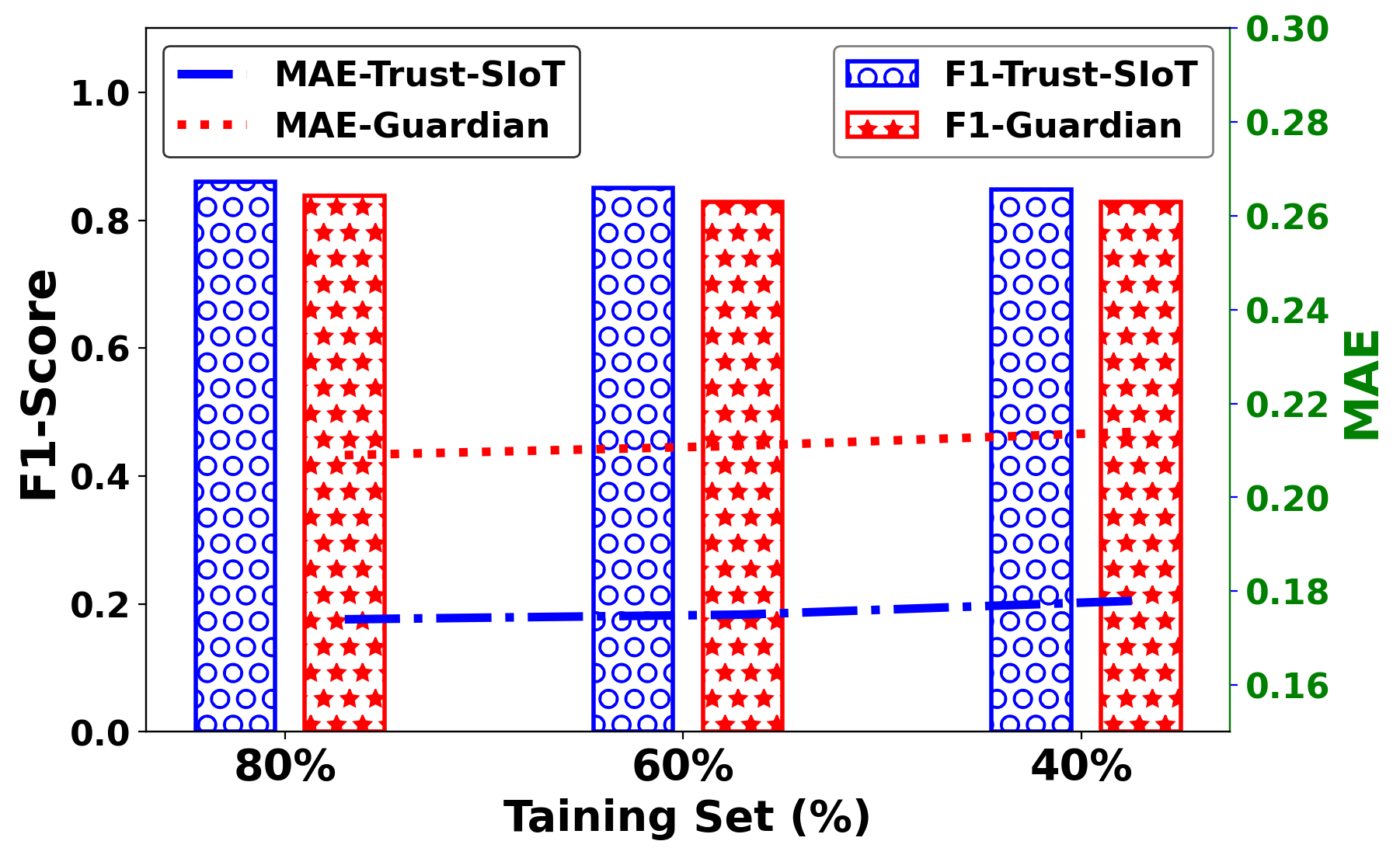}
         \caption{Advogato}
         \label{fig:rob_acc_adv}
     \end{subfigure}%
     \begin{subfigure}[b]{0.25\textwidth}
         \centering
         \includegraphics[width=4.5cm,height=3.5cm]{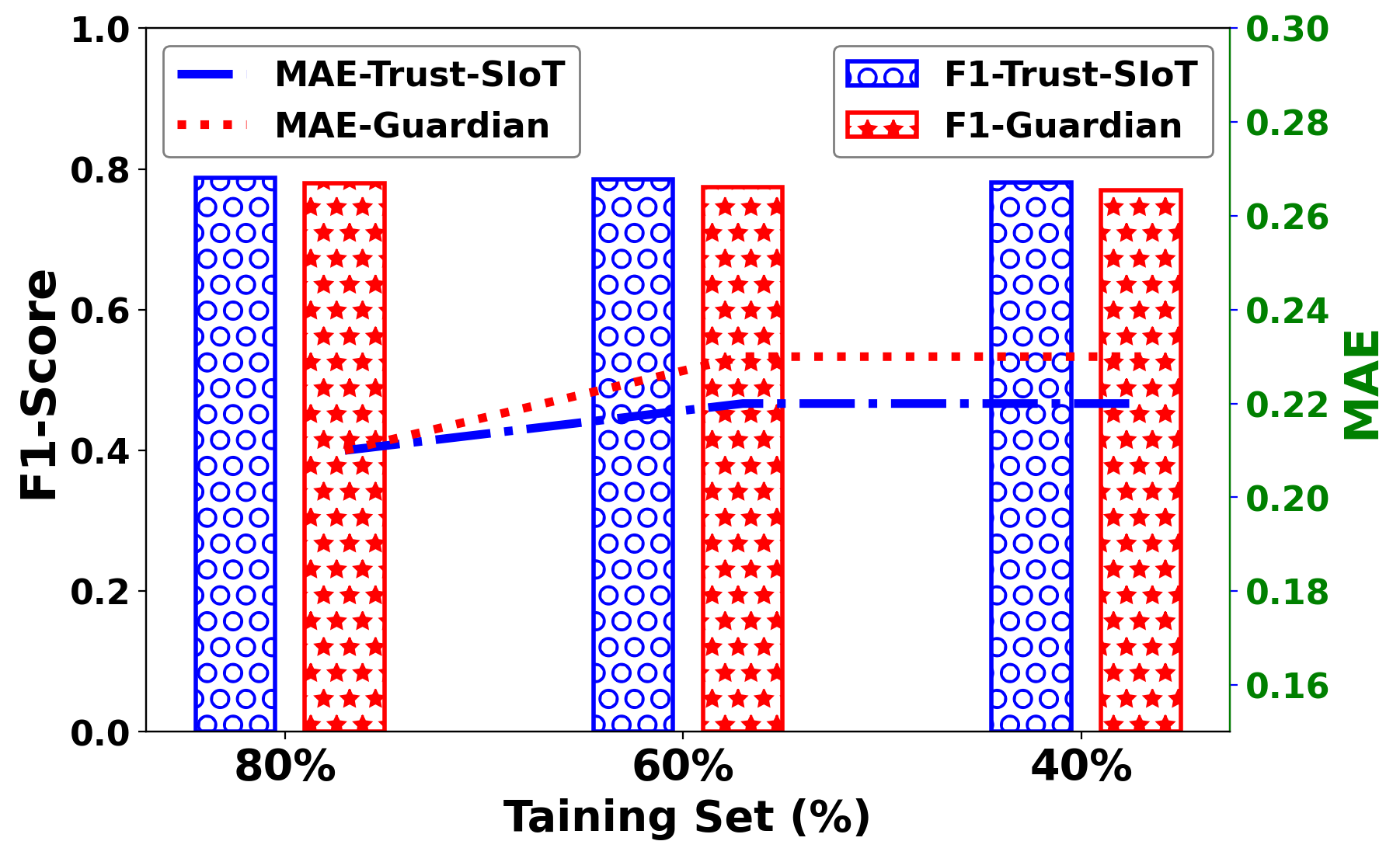}
         \caption{BTC-Alpha}
         \label{fig:rob_acc_btc}
     \end{subfigure}
    \caption{Robustness of \emph{Trust--SIoT} and \emph{Guardian} for Advogato and BTC-Alpha}
\end{figure}

Furthermore, the \emph{robustness} of the proposed approach is evaluated by varying the proportion of training set into $80\%$, $60\%$, and $40\%$ respectively. The reported results for \emph{Advogato} is shown in Table \ref{table:rob_acc_adv} and in Figure \ref{fig:rob_acc_adv}. It can be seen that the \emph{Trust--SIoT} has a negligible performance decrease of $1.1\%$,  $3.3\%$,  $4\%$ in terms of F1-Score, MAE, and MSE respectively when the training set size is reduced to $40\%$. For \emph{Guardian}, the performance evaluation decreases by $1.3\%$, $6.7\%$, $6.2\%$ with regards to 
%F!-score, 
F1-score, MAE, and MSE respectively. Likewise, there is slight decrease in the performance of both \emph{Trust--SIoT} and \emph{Guardian} for \emph{BTC-Alpha} as shown in Table \ref{table:rob_acc_btc} and in Figure \ref{fig:rob_acc_btc}, and there is a similar performance of both the approaches in terms of all the evaluation metrics. 

In essence, the results suggest that the \emph{Trust--SIoT} framework effectively learns the trust relationships between the objects even with the low percentage of the training set, and is able to perform better in comparison with the other approach.  

\begin{table}[h]
    \centering
    \caption{Robustness with varying training set - \emph{Advogato}}
    \begin{tabular}{|c|c|c|c|c|}
    \hline
    $Model$ & $Training \ set $ & $F1-Score$ & $MAE$ & $MSE$\\ \hline \hline
      & 80\% & 0.860 & 0.174 & 0.240 \\
     \emph{Trust--SIoT} & 60\% & 0.857 & 0.175 & 0.248 \\
      & 40\% & 0.850 & 0.180 & 0.250 \\ \hline
     & 80\%& 0.838 & 0.209 & 0.290 \\ 
    \emph{Guardian} & 60\%& 0.830 & 0.212 & 0.290 \\
     & 40\%& 0.827 & 0.223 & 0.302 \\ \hline
   \end{tabular}
    \label{table:rob_acc_adv}
\end{table}

\begin{table}[h]
    \centering
    \caption{Robustness with varying training set - \emph{BTC Alpha}}
    \begin{tabular}{|c|c|c|c|c|}
    \hline
    $Model$ & $Training \ set$ & $F1-Score$ & $MAE$ & $MSE$\\ \hline \hline
      & 80\% & 0.785 & 0.211 & 0.220 \\ 
    \emph{Trust--SIoT}  & 60\% & 0.775 & 0.220 & 0.220 \\
      & 40\% & 0.770 & 0.221 & 0.23 \\ \hline
     & 80\% & 0.781 & 0.215 & 0.219 \\ 
    \emph{Guardian} & 60\% & 0.770 & 0.224 & 0.226 \\
    & 40\% & 0.768 & 0.230 & 0.229 \\ \hline
   \end{tabular}
    \label{table:rob_acc_btc}
\end{table}

\begin{figure*}[!t]
     \centering
     \begin{subfigure}[b]{0.3\textwidth}
         \centering
         \includegraphics[width=\textwidth]{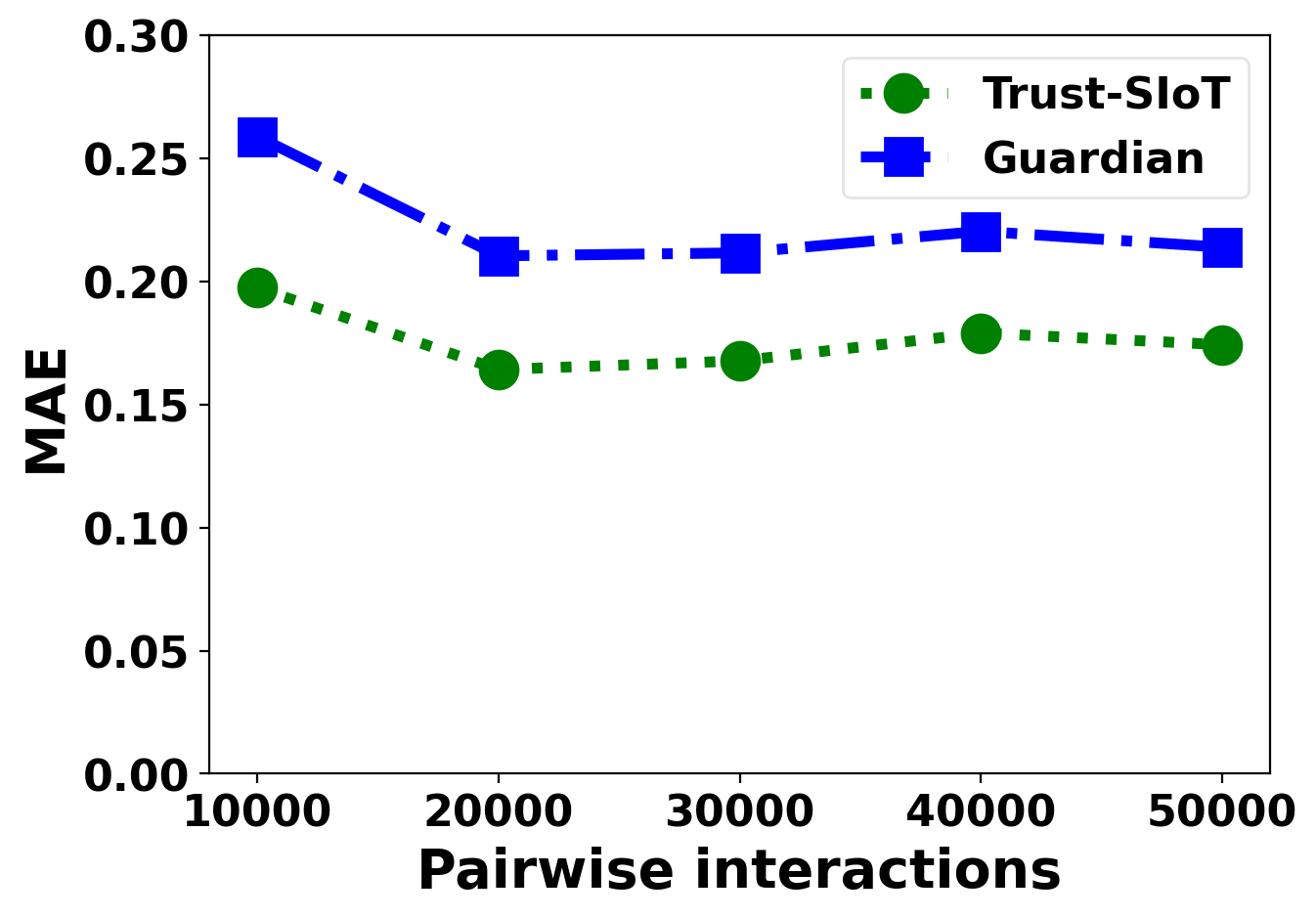}
         \caption{}
         \label{fig:f1_adv}
     \end{subfigure}
     \hspace{0.35em}
     \begin{subfigure}[b]{0.3\textwidth}
         \centering
         \includegraphics[width=\textwidth]{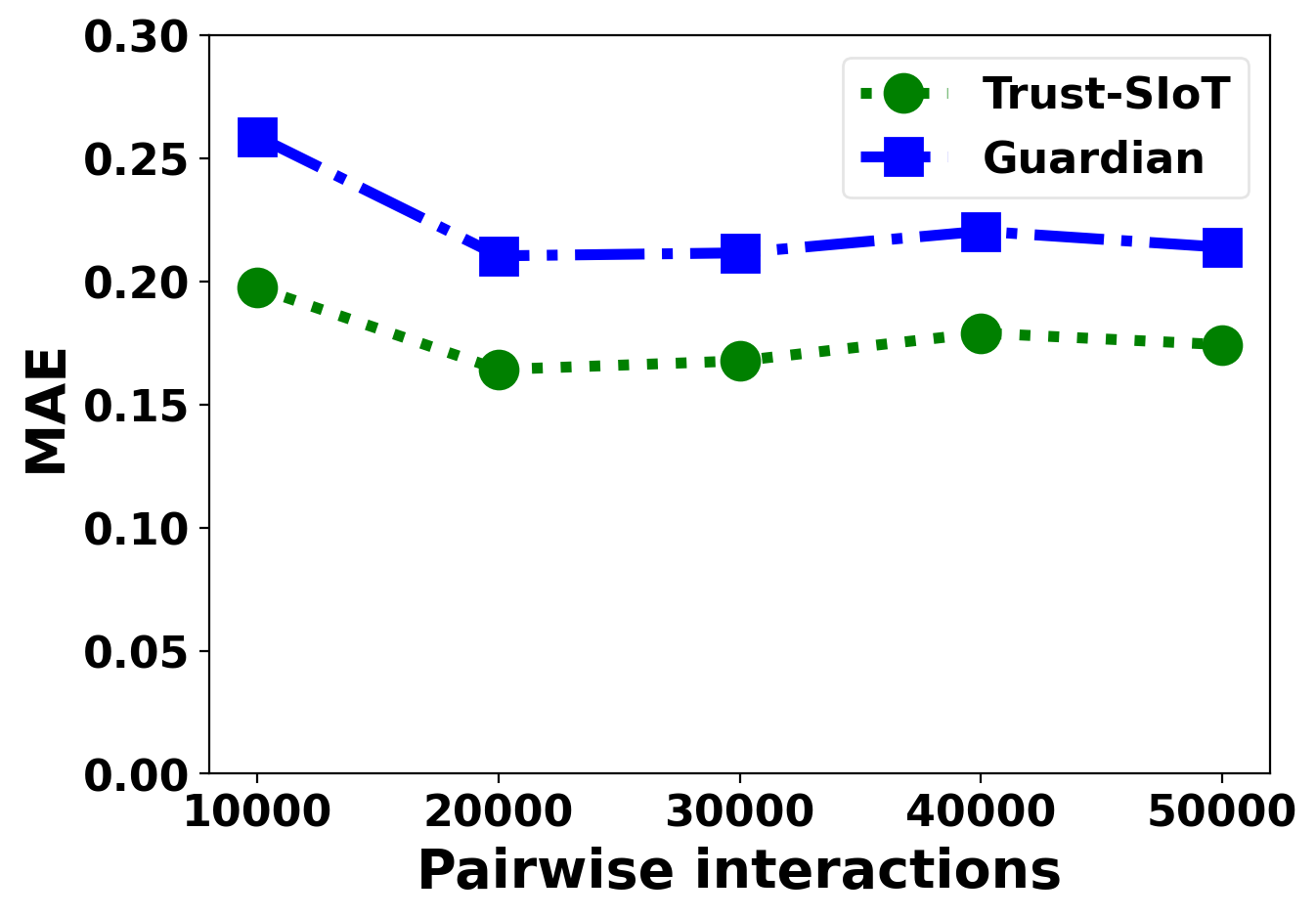}
         \caption{}
         \label{fig:mae_adv}
     \end{subfigure}
     \hspace{0.35em}
     \begin{subfigure}[b]{0.3\textwidth}
         \centering
         \includegraphics[width=\textwidth]{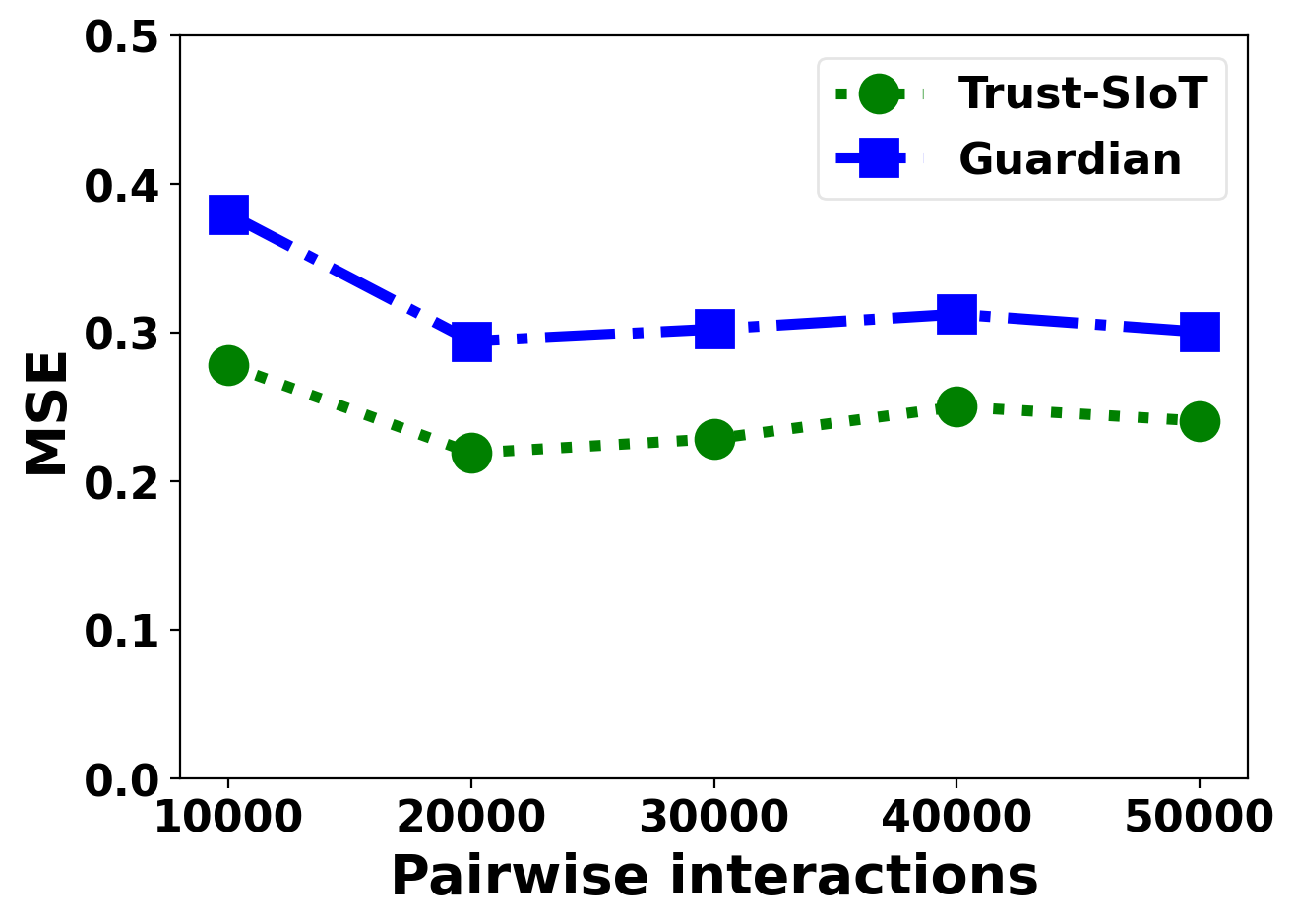}
         \caption{}
         \label{fig:mse_adv}
     \end{subfigure}
        \caption{F1-score, MAE, and MSE with varying pairwise interactions - Advogato}
        \label{fig:scale_adv}
\end{figure*}

We have also evaluated the performance of the proposed framework with the varying number of pairwise interactions to observe the scalability of the \emph{Trust--SIoT}. The scalability of both the approaches for \emph{Advogato} dataset can be seen in Figure \ref{fig:scale_adv}a, \ref{fig:scale_adv}b, and \ref{fig:scale_adv}c in terms of F1-score, MAE and MSE respectively. It is evident from the results that the F1-score of \emph{Trust--SIoT} remains stable with the increase in the pairwise interactions and always remains higher than the \emph{Guardian}. Moreover, the MAE and MSE scores of \emph{Trust--SIoT} remain low over the varying interactions in comparison to \emph{Guardian}. 

% \begin{figure}[b]
% \begin{minipage}[b]{0.5\linewidth}
% \centering
% \includegraphics[scale=0.27]{adv_f1_mae.png}
% \caption{Advogato}
% % \caption{Advogato}
% \label{fig:rob_acc_adv}
% \end{minipage}\hfill\begin{minipage}[b]{0.5\linewidth}
% \centering
% \includegraphics[scale=0.27]{btc_f1_mae.png}
% \captionof{figure}{BTC-Alpha}
% \label{fig:rob_acc_btc}
% \end{minipage}
% \captionsetup{labelformat=empty}
% \caption{\textbf{Figure} \ref{fig:rob_acc_adv} and \ref{fig:rob_acc_btc}: Robustness of \emph{Trust--SIoT} and \emph{Guardian} for Advogato and BTC-Alpha}
% \end{figure}

\begin{figure*}[t]
     \centering
     \begin{subfigure}[b]{0.3\textwidth}
         \centering
         \includegraphics[width=\textwidth]{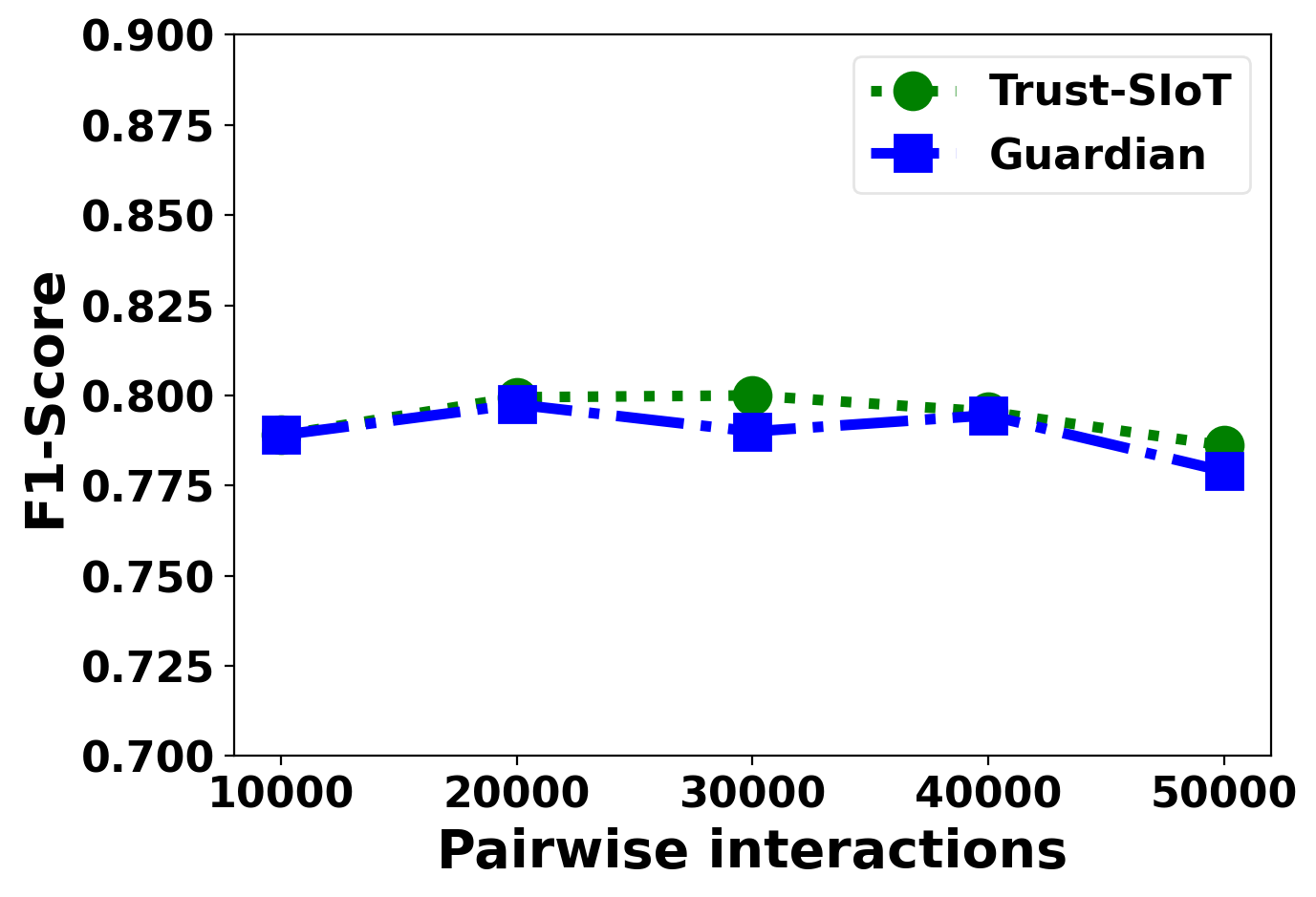}
         \caption{}
         \label{fig:f1_btc}
     \end{subfigure}
     \hspace{0.35em}
     \begin{subfigure}[b]{0.3\textwidth}
         \centering
         \includegraphics[width=\textwidth]{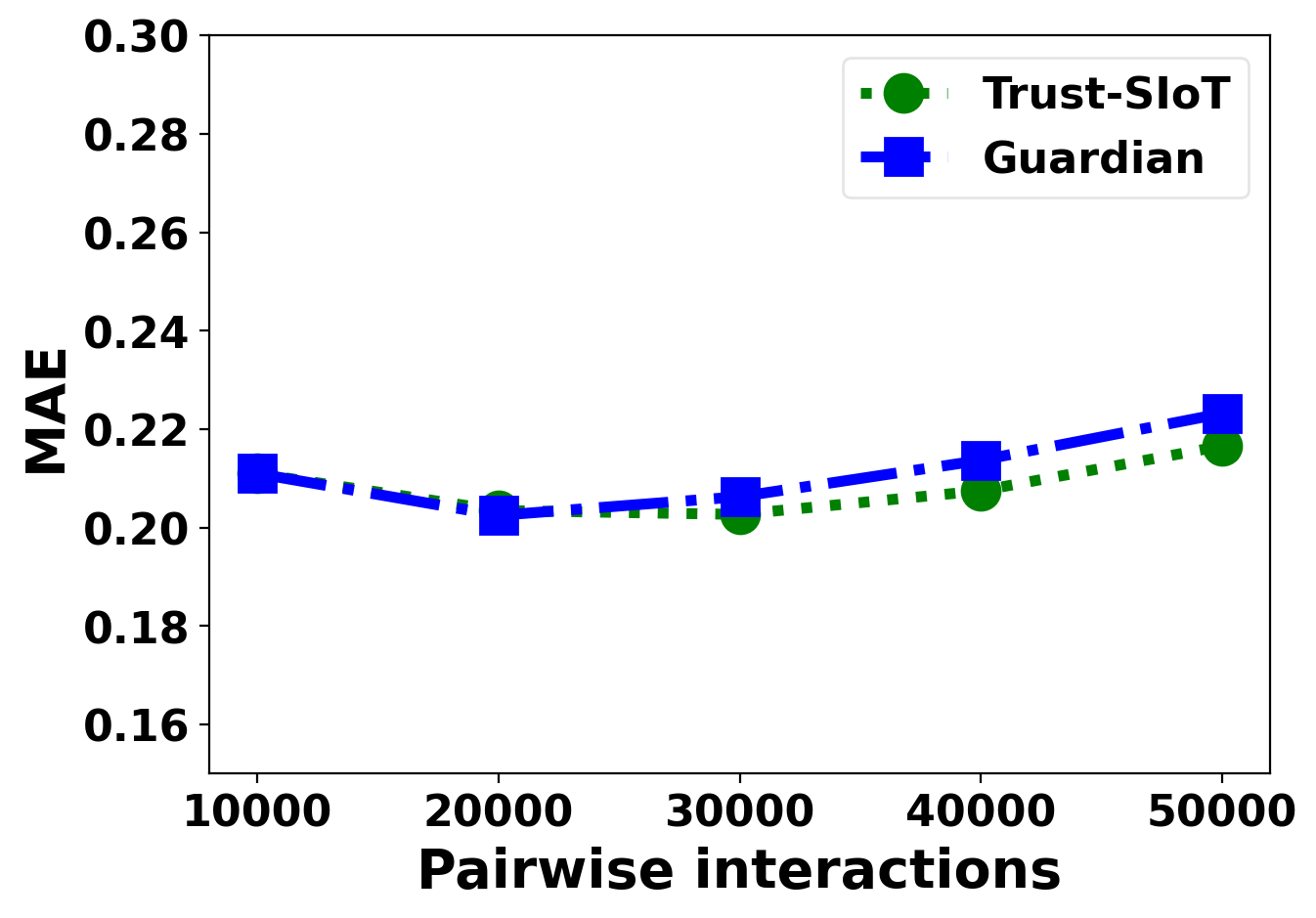}
         \caption{}
         \label{fig:mae_btc}
     \end{subfigure}
     \hspace{0.35em}
     \begin{subfigure}[b]{0.3\textwidth}
         \centering
         \includegraphics[width=\textwidth]{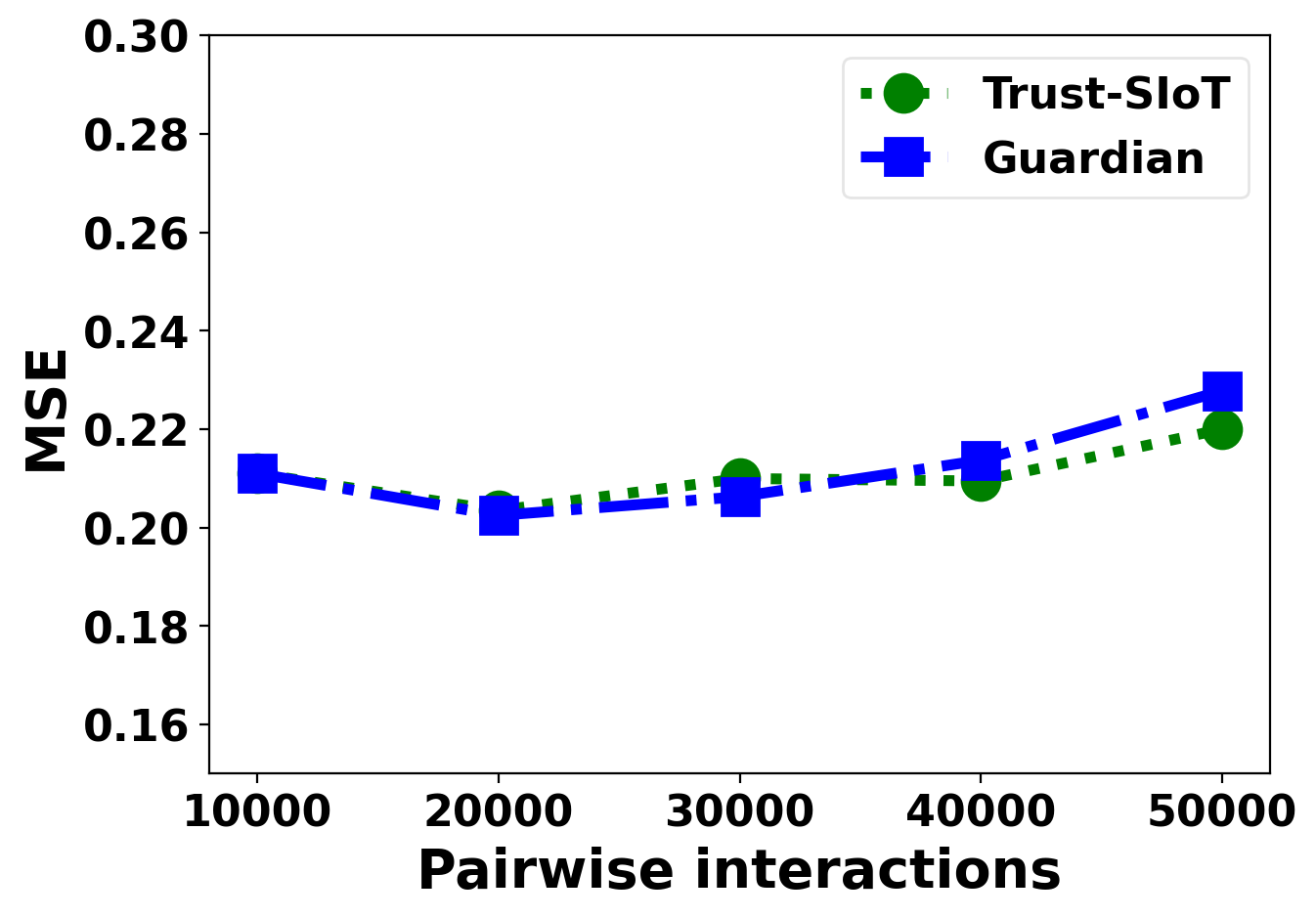}
         \caption{}
         \label{fig:mse_btc}
     \end{subfigure}
        \caption{F1-score, MAE, and MSE with varying pairwise interactions - BTC-Alpha}
        \label{fig:scale_btc}
\end{figure*}

Similarly, Figure \ref{fig:scale_btc}a, \ref{fig:scale_btc}b, and \ref{fig:scale_btc}c present the F1-score, MAE and MSE with varying interactions for \emph{BTC-Alpha} dataset. As shown in Figure \ref{fig:scale_btc}a the F1-score decreases with the increase in number of interaction, however, it is slightly better than \emph{Guardian}. Furthermore, the MAE and MSE scores of both the approaches increase with 
%increases 
growth in number of interactions. Nevertheless, these scores remain low for \emph{Trust--SIoT} in comparison to \emph{Guardian} with the increase in number of interactions. 
In general, it is noteworthy that \emph{Trust--SIoT} performs better on both the datasets and is scalable to generalize for large scale SIoT networks.   

\begin{figure}[t]
\centering
     \begin{subfigure}[b]{0.25\textwidth}
         \centering
         \includegraphics[width=4.5cm,height=3.5cm]{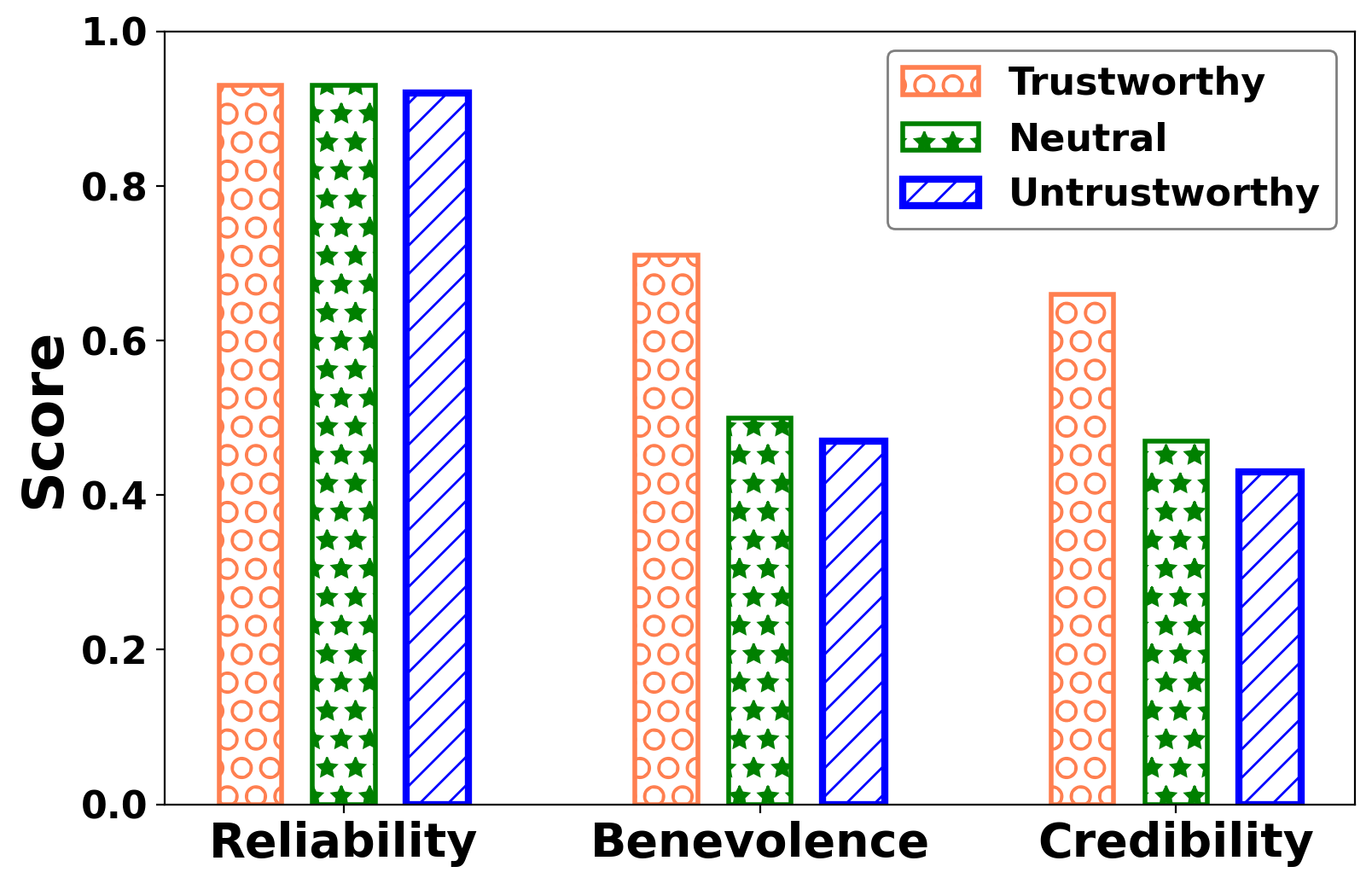}
         \caption{Advogato}
         \label{fig:rbc_adv}
     \end{subfigure}%
     \begin{subfigure}[b]{0.25\textwidth}
         \centering
         \includegraphics[width=4.5cm,height=3.5cm]{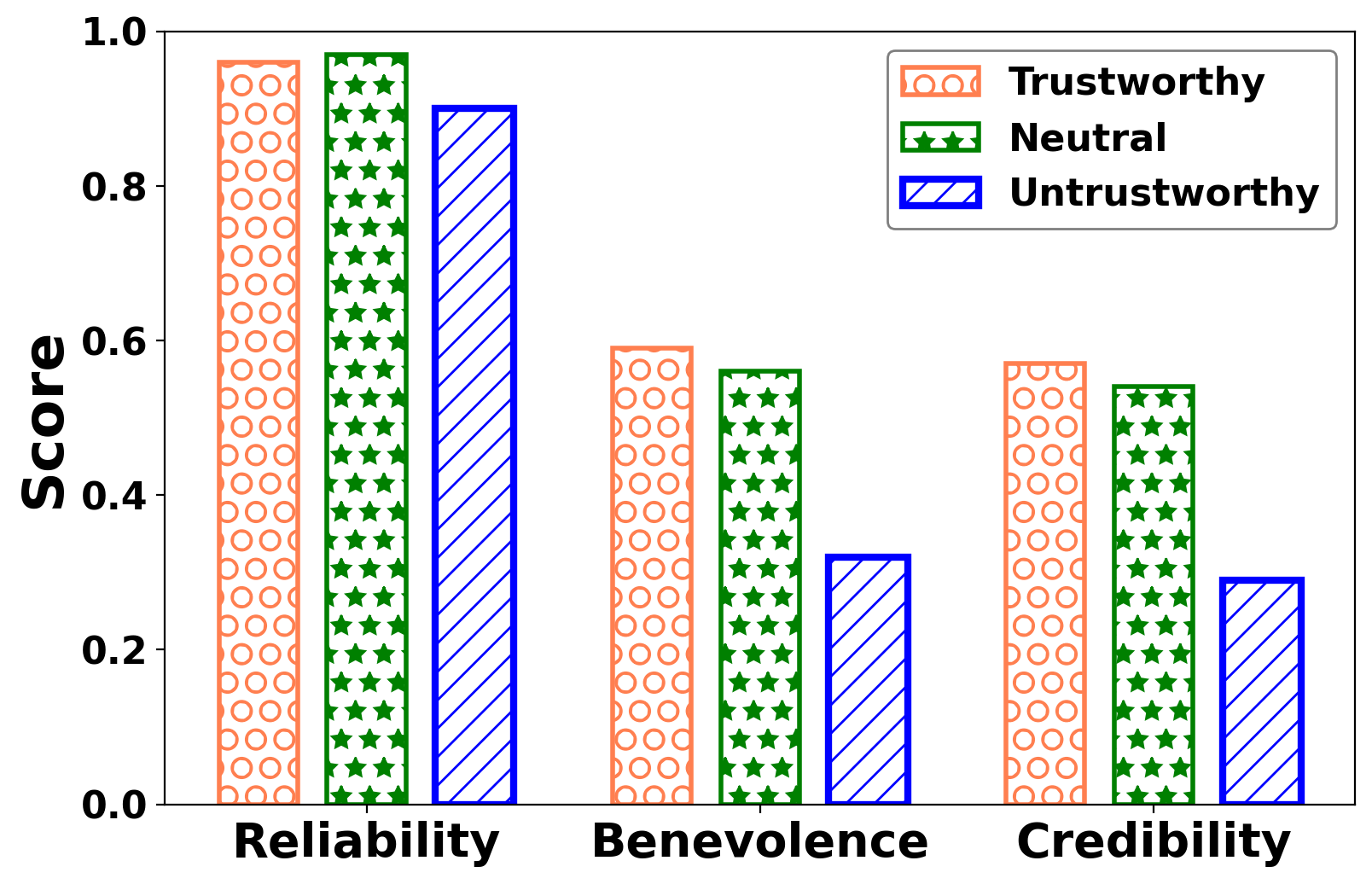}
         \caption{BTC-Alpha}
         \label{fig:rbc_btc}
     \end{subfigure}
    \caption{Dependency analysis of $\mathcal{R}$, $\mathcal{B}$, and $\mathcal{CR}$ for Advogato and BTC-Alpha}
\end{figure}

Finally, we have also analyzed the change in the reliability, benevolence, and credibility of trustworthy, neutral, and untrustworthy objects for both \emph{Advogato} and \emph{BTC-Alpha}. As depicted in Figure \ref{fig:rbc_adv} that for an object to be trustworthy, it needs to be credible and hence object must be reliable and benevolent as the credibility is the integration of both the metrics. Furthermore, it can be observed that the credibility of untrustworthy objects is lower than that of neutral and trustworthy objects. 
Likewise, Figure \ref{fig:rbc_btc} manifests the reliability, benevolence and credibility objects for \emph{BTC-Alpha} datasets. Similar to the results observed in \emph{Advogato}, all three metrics have a high score for trustworthy objects, and these scores are lower for neutral and untrustworthy objects. 

Thus, it is evident that both the reliability and benevolence trust metrics are essential to comment on the credibility of objects in the SIoT network, and the proposed metrics are not dataset 
%depends 
dependant and can be generalized for other large scale SIoT and also for online social networks. 

% \begin{table}[t]
% \begin{minipage}[b]{0.5\linewidth}
% \centering
% \includegraphics[scale=0.27]{adv_rel_ben.png}
% \captionof{figure}{Advogato}
% \label{fig:rbc_adv}
% \end{minipage}\hfill\begin{minipage}[b]{0.5\linewidth}
% \centering
% \includegraphics[scale=0.27]{btc_rel_ben.png}
% \captionof{figure}{BTC-Alpha}
% \label{fig:rbc_btc}
% \end{minipage}
% \captionsetup{labelformat=empty}
% \caption{\textbf{Figure} \ref{fig:rbc_adv} and \ref{fig:rbc_btc}: Dependency Analysis of $\mathcal{R}$, $\mathcal{B}$, and $\mathcal{CR}$ for Advogato and BTC-Alpha}
% \end{table}

\section{Conclusion and Future Work}
\label{sec:conc}

%Adnan: @Subhash - I have now Edited the Conclusion. Please check.

In this paper, we have envisaged an artificial neural network-based trust framework, \emph{Trust--SIoT}, wherein the convergence of dynamic social trust metrics, i.e., direct trust by integrating both current and past interactions, reliability and benevolence, credible recommendations, and the degree of relationships, has enhanced the trustworthiness evaluation of the SIoT objects. The recommendations are ascertained from the credible neighbouring objects and the SIoT knowledge graph is constructed so as to learn the embedding vectors for quantifying the degree of relationships. The experimental analysis suggests that \emph{Trust--SIoT} outperforms the state-of-the-art heuristics, particularly, in terms of the F1, MAE, and MSE measurement. In the near future, we intend to investigate this framework vis-\'a-vis the trust-related attacks in a bid to strengthen its resilience in dynamic SIoT environments.

%In this paper, we have delineated a trustworthy object classification model, (\emph{Trust–-SIoT}). Particularly, we have employed a number of trust metrics explored in both social networks and SIoT characteristics. The convergence of dynamic trust metrics, i.e., data trust metrics, reliability, and benevolence,credible recommendations and degree of relationships have enhanced the trustworthiness evaluation of the SIoT object. Moreover, recommendations are obtained from credible objects from neighbouring objects and the SIoT knowledge graph is constructed to learn the embedding vector to quantify the context-aware degree of relationships. Then, these metrics are fed into an ANN-based heuristic to classify trustworthy objects. Finally, the experimental results show that the proposed model performance better than the existing state-of-the-art with the significant improvement in the value of the F1-score and decrease in the value of MAE and MSE.

%In the future, we plan to further investigate the framework for the individual trust-related attacks to verify the accuracy and resilience in the dynamically changing environment. Furthermore, we plan to apply and explore the trust evaluation techniques employed in the online social network.

\section*{Acknowledgement}
The corresponding author, Subhash Sagar, sincerely acknowledges the generous support of the Higher Education Commission of Pakistan and Macquarie University, Australia for funding the research-at-hand via its ``Macquarie University Research Excellence Award (Allocation No. 2019050)’'. Quan Z. Sheng's work has been partially supported by Australian Research Council Future Fellowship under Grant FT140101247 and 
Discovery Project under Grant DP200102298.
% \begin{table*}
% \centering
% \small
%     \begin{tabular}{c|c|c|c|c|c|c}
%     \hline
%      & \multicolumn{6}{|c}{Trust Frameworks} \\ \cline{2-7}
%     Datasets&\multicolumn{3}{|c|}{\emph{Trust--SIoT}}& \multicolumn{3}{|c}{\emph{Guardian}} \\ \cline{2-7}
%     & F1-Score & MAE & MSE & F1-Score & MAE & MSE \\ \hline
%     BTC-Alpha&29&17&27&20&93&1\\
%     Advogato&19&22&23&20&84&1\\ \hline \hline
%     Improvement & 1&1 &1 &1&1&1 \\ \hline
%     \end{tabular}
%     \caption{Number of respondants for treatment}
%  \end{table*}
 
\bibliographystyle{IEEEtran}
\balance
\bibliography{sagar}

\end{document}